\documentclass[twocolumn]{aastex631}

\usepackage{amsmath, amstext}
\usepackage{apjfonts}

\newcommand{\upenn}{Department of Physics \& Astronomy, University of Pennsylvania, 209 S 33rd St., Philadelphia, PA 19104, USA}
\newcommand{\cca}{Center for Computational Astrophysics, Flatiron Institute, 162 5th Ave, New York, NY 10010, USA}

\newcommand{\half}{\ensuremath{\frac{1}{2}}}

\newcommand{\feh}{\ensuremath{\left[\textrm{Fe/H}\right]}}
\newcommand{\dform}{\ensuremath{d_{\mathrm{form}}}}
\newcommand{\thetaform}{\ensuremath{\theta_{\mathrm{form}}}}
\definecolor{ccolor}{RGB}{53, 56, 255}

\newcommand{\mi}{\textsf{m12i }}
\newcommand{\mm}{\textsf{m12m }}
\newcommand{\mf}{\textsf{m12f }}

\usepackage[section]{placeins}

\shorttitle{New families in our Solar neighborhood}
\shortauthors{Nikakhtar et al.}
\graphicspath{{./}{figures/}}

\begin{document}

\title{New families in our Solar neighborhood:  applying Gaussian Mixture models for objective classification of structures in the Milky Way and in simulations}

\author[0000-0002-3641-4366]{Farnik Nikakhtar}
\email{farnik@sas.upenn.edu}
\affiliation{\upenn}

\author[0000-0003-3939-3297]{Robyn E. Sanderson}
\affiliation{\upenn}
\affiliation{\cca}

\author{Andrew Wetzel}
\affiliation{Department of Physics \& Astronomy, University of California, Davis, CA 95616, USA}

\author{Sarah Loebman}
\affiliation{Department of Physics \& Astronomy, University of California, Davis, CA 95616, USA}
\affiliation{ Department of Physics, University of California Merced, 5200 North Lake Rd., Merced, CA, 95343, USA}

\author[0000-0002-0920-809X]{Sanjib Sharma}
\affiliation{Sydney Institute for Astronomy, School of Physics, A28, The University of Sydney, NSW 2006, Australia}
\affiliation{Center of Excellence for Astrophysics in Three Dimensions (ASTRO-3D), Australia}

\author{Rachael Beaton}
\affiliation{Department of Astrophysical Sciences, Princeton University, 4 Ivy Lane, Princeton, NJ 08544}
\affiliation{The Observatories of the Carnegie Institution for Science, 813 Santa Barbara St., Pasadena, CA 91101, USA}

\author{J. Ted Mackereth}\thanks{Banting Fellow}
\affiliation{Canadian Institute for Theoretical Astrophysics, University of Toronto, 60 St. George Street, Toronto, ON, M5S 3H8, Canada}
\affiliation{Dunlap Institute for Astronomy and Astrophysics, University of Toronto, 50 St. George Street, Toronto, ON M5S 3H4, Canada}

\author{Vijith Jacob Poovelil}
\affiliation{Department of Physics \& Astronomy, University of Utah, Salt Lake City, UT, 84112, USA}

\author{Gail Zasowski}
\affiliation{Department of Physics \& Astronomy, University of Utah, Salt Lake City, UT, 84112, USA}

\author{Ana Bonaca}
\affiliation{Center for Astrophysics, Harvard \& Smithsonian, 60 Garden Street, Cambridge, MA 02138, USA}

\author{Sarah Martell}
\affiliation{School of Physics, UNSW, Sydney, NSW 2052, Australia}
\affiliation{Center of Excellence for Astrophysics in Three Dimensions (ASTRO-3D), Australia}

\author[0000-0002-4912-8609]{Henrik J\"onsson}
\affiliation{Materials Science and Applied Mathematics, Malm\"o University, SE-205 06 Malm\"o, Sweden}

\author{Claude-Andr{\'e} Faucher-Gigu{\`e}re}
\affiliation{Department of Physics and Astronomy and CIERA, Northwestern University, 2145 Sheridan Road, Evanston, IL 60208, USA}



\begin{abstract}

The standard picture of galaxy formation motivates the decomposition of the Milky Way into 3--4 stellar populations with distinct kinematic and elemental abundance distributions: the thin disk, thick disk, bulge, and stellar halo. To test this idea, we construct a Gaussian mixture model (GMM) for both simulated and observed stars in the Solar neighborhood, using measured velocities and iron abundances (i.e., an augmented Toomre diagram) as the distributions to be decomposed. We compare results for the Gaia-APOGEE DR16 crossmatch catalog of the Solar neighborhood with those from a suite of synthetic Gaia-APOGEE crossmatches constructed from FIRE-2 cosmological simulations of Milky Way-mass galaxies. We find that in both the synthetic and real data, the best-fit GMM uses \emph{five} independent components, some of whose properties resemble the standard populations predicted by galaxy formation theory. Two components can be identified unambiguously as the thin disk and another as the halo. However, instead of a single counterpart to the thick disk, there are three intermediate components with different age and alpha abundance distributions (although these data are not used to construct the model). We use decompositions of the synthetic data to show that the classified components indeed correspond to stars with different origins. By analogy with the simulated data, we show that our mixture model of the real Gaia-APOGEE crossmatch distinguishes the following components: (1) a classic thin disk of young stars on circular orbits (46\%), (2) thin disk stars heated by interactions with satellites (22\%), (3, 4) two components representing the velocity asymmetry of the alpha-enhanced thick disk (27\%), and (5) a stellar halo consistent with early, massive accretion (4\%).

\end{abstract}



\section{Introduction}
\label{sec:intro}

The kinematics and elemental  abundances of the Milky Way's stars are thought to contain clues to the formation history of the Galaxy we live in \citep[e.g.][]{2002ARA&A..40..487F}. In the classic picture, the distribution of stars in velocity and elemental  abundances has a relatively small number of distinct components linked to different formation epochs:
\begin{itemize}
\item At very early times, star formation and proto galaxy merging take place in a relatively chaotic environment, leading to a roughly spheroidal distribution variously referred to as an ``early spheroid'' \citep[e.g.][]{elmegreen2008}. Stars formed very fast in this epoch, so despite starting from gas almost free of metals, the resultant population is quite metal-rich.

\item A subsequent epoch of accretion creates a hot disk structure (still relatively metal-poor), which forms stars in the present-day \emph{thick disk} \citep[e.g.][]{forbes2012,bird2013} and/or stars formed early on in a thin disk are heated by scattering processes and radial migration to form the thick disk \citep{2020arXiv200503646S, 2009MNRAS.396..203S}.

\item The \emph{thin disk} is formed by colder and more gradual accretion of more metal-rich gas, regulated by feedback from relatively steady star formation, in a process that continues to the present day \citep[e.g.,][]{brook2012, 2017arXiv171203966G, 2017MNRAS.467.2430M, Stern2020}.

\item Accretion of smaller, more metal-poor components contributes an additional, roughly spheroidal with a larger scale radius but with substructures, commonly referred to as the ``outer halo'' or ``\emph{accreted halo}'' \citep[e.g.][]{searlezinn}. \footnote{These two terms, though sometimes used interchangeably, are not synonymous: accreted material, especially from early epochs, can certainly be found at small radii while stars formed in outflows from regions of high star formation in the disk \citep{2020MNRAS.494.1539Y}, and those kicked out by interactions with satellite galaxies \citep{2018MNRAS.481..286L}, can reach large radii \citep{2018MNRAS.480..652E, 2017MNRAS.465.2212S}} In studies of the Solar neighborhood, this component and the first one are often jointly referred to as the ``halo'', which is a suitable simplification given the short scale radius and complex kinematics of the Milky Way's bulge-like component \citep{2012ApJ...744L...8G}.
\end{itemize}

This picture sets up the expectation of a multi-component stellar distribution in the Solar neighborhood, with old, spheroidally distributed stars at the lowest metallicities (the ``halo''); young, metal-rich stars in a kinematically cold ``thin disk"; and a population intermediate in age, metallicity, and kinematics, commonly referred to as the ``thick disk''. Sustained star formation during the cooling of the gas reservoir, as well as gradual dynamical heating of the resulting stellar population, would predict a smooth correspondence between kinematic temperature, metallicity and age in the disk, with kinematically hotter stars being older and more metal-poor.

Many papers prior to this one have sought to test these ideas and refine our understanding of the processes that built the Solar neighborhood by selecting stars based on their kinematics, and studying their abundance distributions \citep[e.g.,][]{bensby2003, venn2004, 2013ApJ...771...67I, nissenschuster, bonaca17, DeokkeunBeers2020, Hayden2020}. Most often, these kinematic cuts are performed in the Toomre diagram, where the $x$ axis has the velocity of stars in the direction of Galactic rotation $V_Y$, and the $y$ axis has the perpendicular component $V_{XZ} \equiv \sqrt{V_X^2 + V_Z^2}$. Here, $V_X$ is along the Sun-Galactic center direction, and $V_Z$ is perpendicular to the disk plane in the direction of the total angular momentum.

\begin{figure}
\includegraphics[width=\columnwidth]{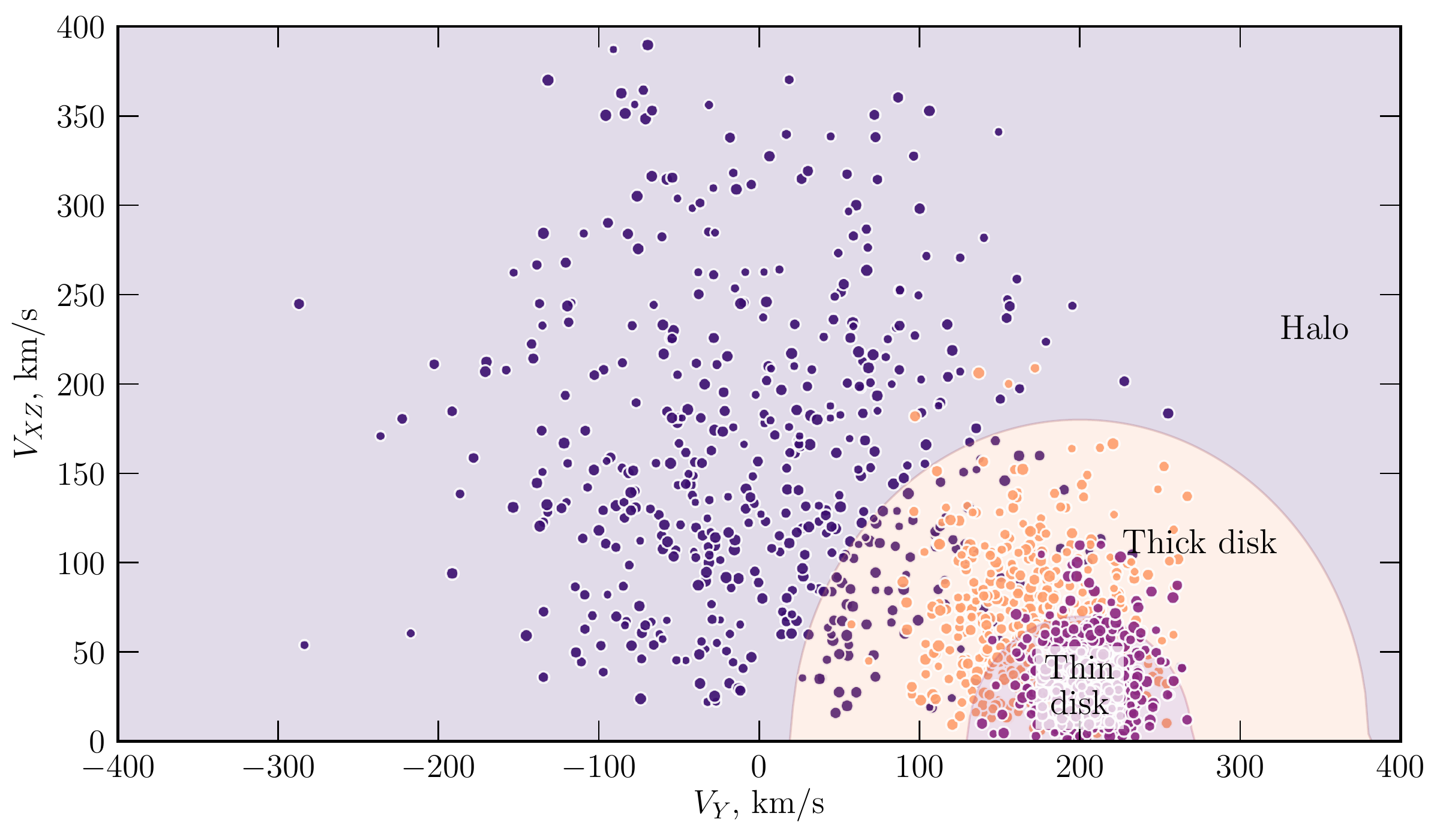}
\caption{Typical regions of the Toomre diagram based on observation in the Solar neighborhood and ascribed to different chemokinematic components following \citet{bensby2003}: the metal-poor ``halo'' (purple), the intermediate ``thick disk'' (orange), and the metal-rich ``thin disk'' (magenta).}
\label{fig:classicToomre}
\end{figure}

Figure~\ref{fig:classicToomre} shows the Toomre diagram of stars drawn from traditional components of the Galaxy measured by \citet{bensby2003}, with the thin disk in magenta, the thick disk in orange and the halo in purple. These components are overlapping, but a selection criterion based on the relative velocity with respect to the Local Standard of Rest (LSR) can preferentially select thin disk stars comoving with the LSR (magenta shaded region in Figure~\ref{fig:classicToomre}), thick disk stars at intermediate distance from the LSR (orange shaded region) and halo stars moving at high velocity with respect to the LSR (purple shaded region).

In an alternative to such kinematic selections, other studies have selected disk stars based on their elemental  abundances, and from their spatial distributions tested the idea of a continuous transition between thin and thick disks \citep[e.g.,][]{bovy2012a,bovy2012b,bovy2016,Mackereth2017a}. Some cosmological simulations \citep[e.g.,][]{ma2017, bird2020} also do not predict a clean/sharp transition from the thick to the thin disk, but rather, a more gradual settling of the stellar disk (note, however, that some simulations such as the FIRE simulations analyzed in this paper suggest a sharper transition in the properties of the gas disk, as galaxies transition from highly bursty to more steady star formation rates, \citep[e.g.,][]{Stern2020}). Still others were interested in searching for local interlopers from the halo to assess the Milky Way's accretion history, simultaneously employing both kinematic and metallicity cuts to select this relatively small population from the overwhelmingly more numerous disk stars \citep[e.g.,][]{helmi2017, herzogarbeitman}. 

The assumptions these approaches make about links between the kinematics and metallicity and/or alpha-enhancement of stars have begun to be challenged with the advent of \emph{Gaia}'s exquisite kinematic information \citep{gaia.mission}. For example, several authors have pointed out the presence of a population that is either an intermediate component between the thick disk and the halo \citep{bonaca17, posti2017, Belokurov2018} or a sense of rotation in the halo itself \citep{deason2017, kafle2017}. The Gaia-Enceladus or ``Sausage`` structure \citep{Belokurov2018, helmi18, Myeong18, 2018MNRAS.477.5072M} is likely a massive contributor to the local neighborhood and has 30\%--50\% of the halo stellar mass \citep{Mackereth2020}.

This influx of new information both enables and motivates the relaxation of some of the assumptions about the structure of the Solar neighborhood, in favor of allowing the data itself to tell us what the distribution looks like. In this work, we take the agnostic approach of modeling the stellar distribution as a mixture of Gaussians, with the goal of imposing as few assumptions as possible on its observed quantities.

We construct a mixture model (described briefly in \S\ref{sec:gmm}) of the velocities and iron abundances of stars in the Solar neighborhood, leaving the number of Gaussian components in the model free to vary and using an information criterion to pick the most suitable number. We test this approach on a new set of mock Gaia-APOGEE catalogs generated from FIRE-2 cosmological simulations\footnote{See the FIRE project website: http://fire.northwestern.edu} of a Milky Way-mass galaxy \citep{Sanderson19.firegaiadr2} (described in \S\ref{sec:apogee}). From these mock catalogs, we find that the best-fit model is consistent with previous arguments on the origins of Solar neighborhood stars and their spatial, kinematic, and abundance distributions (\S\ref{sec:mock}). Perhaps surprisingly, we find that the optimal decomposition features five components in all the simulations: two analogous to the thin disk and the halo, but instead of a single counterpart to the thick disk, there are three intermediate components with distinct age-alpha distributions and formation histories, even though these data are not used to fit the model. We then apply the same strategy to stars in the Solar neighborhood using the Gaia DR2 catalog crossmatched with the APOGEE DR16 (\S\ref{sec:apogee}) survey  and find a similar result (\S\ref{sec:real}). We conclude in \S\ref{sec:discussion} by drawing analogies with the simulated surveys to postulate distinct origins for the five components identified in our Milky Way.

\section{Gaussian mixture modeling}
\label{sec:gmm}

A Gaussian mixture model (GMM) describes a distribution of $n_s$ data points (\emph{samples}) $\vec{x}_j$ using a combination of $n_c$ Gaussian distributions with independent mean values $\vec{\mu}^i$ and covariance matrices $\mathbf{\Sigma}^i$.
The $\vec{x}_j$ contain the $n_f$ \emph{features} (i.e. dimensions of data) used to determine the probabilities $p$ that each of the data points belongs to each of the $n_c$ Gaussian components. 

Thus for a given sample,
\begin{equation}
p(\vec{x}_j | \vec{\tau}, \{\vec{\mu}\}, \{\Sigma\}) = \sum_{i=1}^{n_c} \tau_i f_i(\vec{x}_j |\vec{\mu}^i,\Sigma^i),
\end{equation}
where $f_i$ is a $n_f$-dimensional normal distribution with the given mean and covariance, and $\tau_i$ is the relative weight of each Gaussian subject to the constraint that $\sum_{i=1}^{n_c} \tau_i = 1$.

The GMM is thus specified by $n_c - 1$ free weight parameters $\vec{\tau}$, the $n_c n_f$ means $\{\vec{\mu}\}$, and the $n_c n_f (n_f+1)/2$ components of the positive-definite, symmetric covariance matrices $\{\Sigma\}$, for a total of 
\begin{equation}
\label{eq:npars}
n_p = \frac{n_c}{2} (n_f^2 + 3n_f +2) - 1
\end{equation}
free parameters in the model. The probability of data point $j$ belonging to component $k$ in the GMM, also known as the \emph{responsibility} $\mathfrak{R}$, is
\begin{equation}
\mathfrak{R}_k(\vec{x}_j | \vec{\tau}, \{\vec{\mu}\}, \{\Sigma\}) = \frac{\tau_k f_k(\vec{x}_j |\vec{\mu}^k,\Sigma^k)}{\sum_{i=1}^{n_c} \tau_i f_i(\vec{x}_j |\vec{\mu}^i,\Sigma^i)}.
\end{equation}
The model assigns a \emph{label} to each sample, i.e. identifies the Gaussian component to which the data point with coordinates $\vec{x}_j$ is most likely to belong, by choosing the component with the highest $\mathfrak{R}_k$. Thus the GMM acts as both a description of the overall density distribution of the $n_s$ samples in $n_f$-dimensional space, and as an unsupervised classifier that places each sample into one of $n_c$ groups. We use the implementation provided in the Python package {\tt scikit-learn} \citep{scikit-learn}.

In our case the features used for classification will be the three-dimensional velocities $\{V_X, V_Y, V_Z \}$ of the $n_s$ stars, where $V_Y$ is the Galactocentric velocity in the direction of the disk rotation, $V_X$ along the Sun--Galactic center direction, and $V_Z$ perpendicular to the disk plane in the direction of the total angular momentum. We add the iron abundances \feh\ as a fourth dimension or feature, so for our case $n_f=4$.

The reduction of dimensionality in moving from a three-dimensional velocity vector to two Toomre components of velocity $\{V_Y, V_{XZ}\}$ is motivated by the underlying symmetry of the Galaxy, but since there is a zero cutoff in $V_{XZ}$ and our model is a mixture of Gaussians, we construct our model with the three-dimensional velocity vector plus metallicity and just represent obtained clusters in the Toomre sub-space plus metallicity. Likewise, we use Cartesian coordinates rather than cylindrical or spherical coordinates for the velocities to avoid imprinting assumptions about symmetries. Surveys of the current generation are for the most part embedded in the Solar neighborhood, where the assumption of axisymmetry is appropriate. Symmetry assumptions like this are less appropriate to future surveys exploring a larger volume of the Galaxy \citep[e.g.][]{2019ApJ...883..103B}, and the framework and intuition developed in this work lay ground for this transition.

To find the best-fit GMM, we start by choosing a number of components $n_c$. We initialize their means, covariances, and weights by preliminarily labeling each sample using $k$-means clustering \citep{kmeans}, and maximizing the likelihood
\begin{eqnarray}
\ln \mathcal{L}(\vec{x} | \vec{\tau}, \{\vec{\mu}\}, \{\Sigma\}) &=& \sum_{j=1}^{n_s} \sum_{i=1}^{n_c} \left[ \log \tau_i - \half \log |\Sigma^i| - \frac{n_f \log(2 \pi)}{2} \right. \nonumber \\
&& \hspace{-1cm}   \left. - \half \left( \vec{x}_j - \vec{\mu}^i \right)^T 
\cdot (\Sigma^i)^{-1} \cdot \left(\vec{x}_j - \vec{\mu}^i \right) \right], \label{eq:lik}
\end{eqnarray}
using expectation-maximization \citep{empaper}, which determines the best-fit weights, means, and covariances. We then repeat this process for different values of $n_c$ and determine the number of components that minimizes the Bayes information criterion \citep{schwarz1978},
\begin{equation}
BIC = - 2 \ln \mathcal{\hat{L}} +  n_p\ln(n_s),
\end{equation}
where $\mathcal{\hat{L}}$ is the maximum value of the likelihood function given by Equation \eqref{eq:lik} and $n_p$ is the total number of free parameters in the model, given by Equation \eqref{eq:npars}. This criterion compares the maximum likelihood values for different numbers of components (the first term) while including a penalty for introducing additional parameters into the model (the second term), to account for the fact that a model with more free parameters will always produce a better fit. The value of the penalty is derived from an asymptotic expansion of the Bayes evidence as the sample size approaches infinity, under the assumption that the data are independent samples from a distribution with an exponential form (such as a Gaussian). This agrees with the fundamental assumption of the GMM, which motivates our use of this criterion for model selection rather than an information-theory-based criterion such as the Akaike information criterion \citep[AIC;][]{akaike}. The BIC's penalty for adding model parameters, weighted by $\ln n_s$, strongly prefers models with lower $n_p$ relative to the AIC \citep{schwarz1978}.

The assumption that the data we are fitting are truly drawn from a combination of Gaussian components is not necessarily a great one; in fact, it is demonstrably not true for the Toomre coordinates, as we will discuss further in (\S\ref{sec:mock}), so the penalty in the BIC for adding extra components to the mixture model is at best an approximation. Thus although in the idealized case one would look for the minimum BIC value to select the preferred number of components in the model, in practice we do so by increasing $n_c$ just until the BIC stops rapidly decreasing, which is called the \textit{Elbow} rule/method \citep{elbow}. 

\section{Observational and Mock Gaia-APOGEE Catalogs}
\label{sec:apogee}

To apply the concept of separating populations of stars using mixture modeling, we created a suite of mock catalogs mimicking the crossmatch between the Gaia Data Release 2 \citep{gaia.mission, gaia.dr2} and the 16th data release (DR16) of the Apache Point Observatory Galactic Evolution Experiment (APOGEE) \citep{apogee.dr16, sdss.dr16}.

\subsection{Observed Catalog}
\label{subsec:obsCat}

APOGEE-2 \citep{apogee.Majewski2017} is a dual hemisphere survey that uses cloned spectrographs \citep{wilson_2019} operating each at the Apache Point Observatory on the Sloan Foundation 2.5m telescope \citep{gunn_2006} and at Las Campanas Observatory on the duPont Telescope \citep{Bowen_1973}. APOGEE targets primarily red giant stars in all components of the Milky Way, with substantial additional numbers of main sequence and massive evolved stars, which are selected using a simple set of dereddened-color and magnitude criteria. \citep[][Beaton et al.~(in prep.), Santana et al.~(in prep.) ]{zasowski_2013,zasowski_2017}. The exact criteria vary by location in the Galaxy and the length of time a given field will be observed. 

A custom processing pipeline \citep{nidever_2015} reduces the data and calculates heliocentric radial velocities (RVs), and the APOGEE Stellar Parameters and Abundances Pipeline \citep[ASPCAP,][]{garciaperez_2016} produces fundamental stellar parameters (e.g., $\log(g)$, $T_{\rm eff}$) and elemental abundances for up to 26 species. The DR16 catalog contains measurements for ~430,000 stars. We also make some quality selections on the data to remove stars with very small ``observed'' parallaxes (i.e. spuriously large distances, $\pi < 0.1 \mu$as or $d>100$ kpc) and/or large measurement errors on the parallax ($\Delta\pi / \pi > 0.1$), metallicity and radial velocity. After applying these quality cuts around ~150,000 stars remained, which give us a radial coverage of ~4--12 kpc Galactocentric distances.

\subsection{Mock Catalog}
\label{subsec:mockCatalog}

To create the mock Gaia-APOGEE crossmatches, we start from the synthetic Gaia surveys \citet{Sanderson19.firegaiadr2} created from 3 Milky Way-mass galaxies in the Latte suite (first introduced in \cite{Wetzel2016}) of FIRE-2 cosmological simulations \citep{Hopkins2018}, which feature self-consistent clustering of star formation in dense molecular clouds and thin stellar/gaseous disks in live cosmological halos with satellite dwarf galaxies and stellar halos. In each of those simulations, there are 3 solar viewpoints that generate nine synthetic Gaia-like surveys (3 galaxies $\times$ 3 neighborhoods). The synthetic stars are sampled by assuming each star particle represents a single stellar population. The simulations have initial star particle masses of 7070 $M_\odot$, but because of stellar mass loss, a typical star particle, at $z$ = 0, has a mass of $\approx$ 5000 $M_\odot$. At each neighborhood, dust extinction is computed from the simulated gas metallicity distribution.

Regarding the iron abundance, it is important to bear in mind that supernovae (core-collapse and Ia) and stellar winds generate and disperse metals, which are then deposited into gas particles. For supernovae Ia, the stellar nucleosynthesis yields are adopted from \cite{Iwamoto1999}, where the rates follow \cite{Mannucci2006}, including both prompt and delayed populations. For core-collapse supernovae, yields are from \citet{Nomoto2006}; for stellar winds (AGB and O/B-stars), yields are from a compilation of \citet{VanDenHoek1997, Marigo2001, Izzard2004}.

These simulations also include an explicit treatment for unresolved turbulent diffusion of metals in gas, which produces more realistic abundance distributions in both the MW-like galaxies and in their satellite dwarf galaxies \citep{Su2017, Hopkins2018, Escala2018}.

We also add an APOGEE-like error model for the elemental abundances [Fe/H], [Mg/Fe], [C/Fe], [N/Fe], [S/Fe], [O/Fe], [Si/Fe], and [Ca/Fe] as determined in Poovelil et al. (in prep) \footnote{A similar process was ultimately adopted for DR16 as is described in \citet{apogee.dr16}.} where we assume S/N of 100 for every star. Although we only use [Fe/H] for constructing the GMM, we need to track individual abundances (which are not used to fit the model) to see the physical origin and formation histories of each component. In this paper, the [Mg/Fe] is used for this purpose, but other abundances will be explored in forthcoming works.

\begin{figure*}
\begin{center}
\includegraphics[width=\columnwidth]{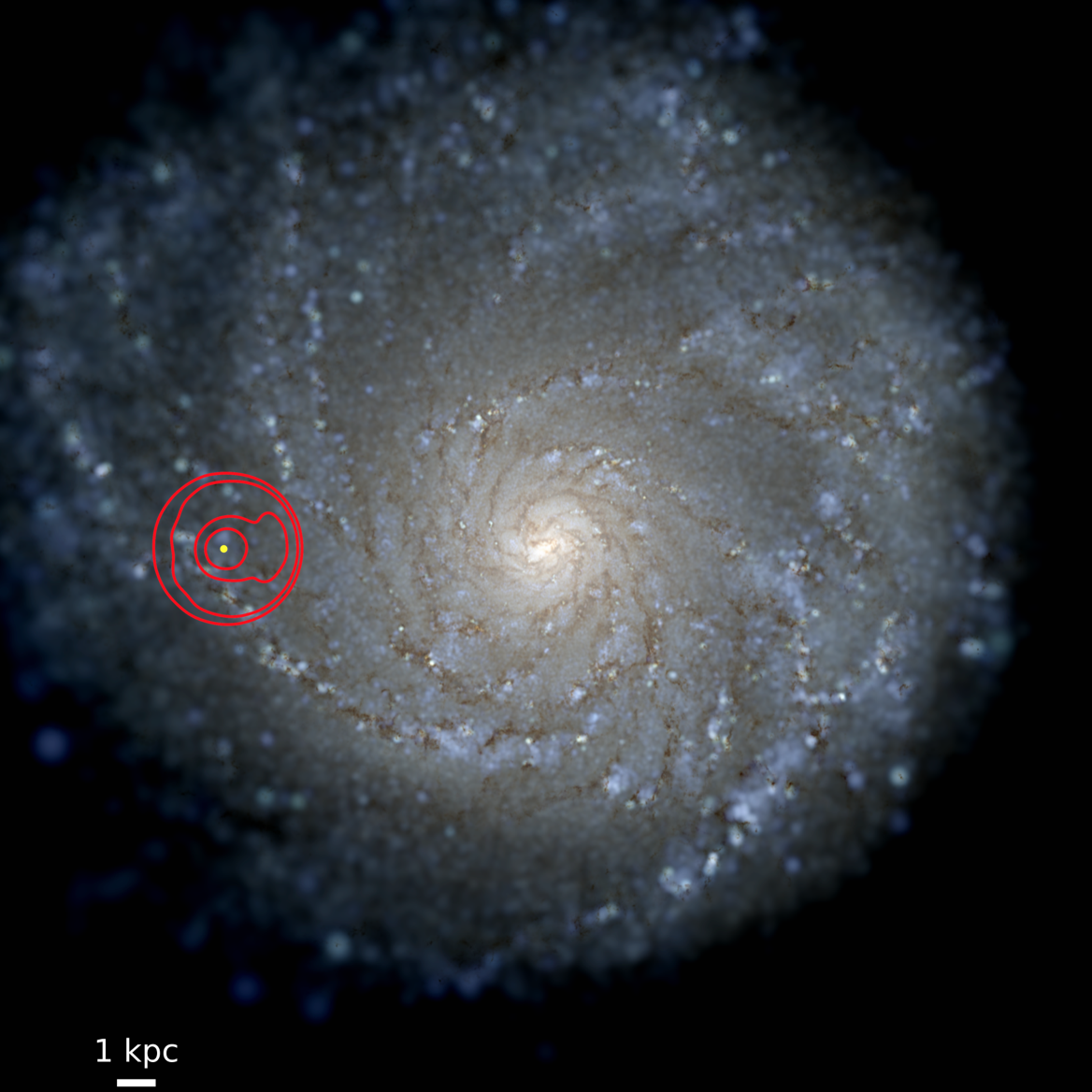}
\quad
\includegraphics[width=\columnwidth]{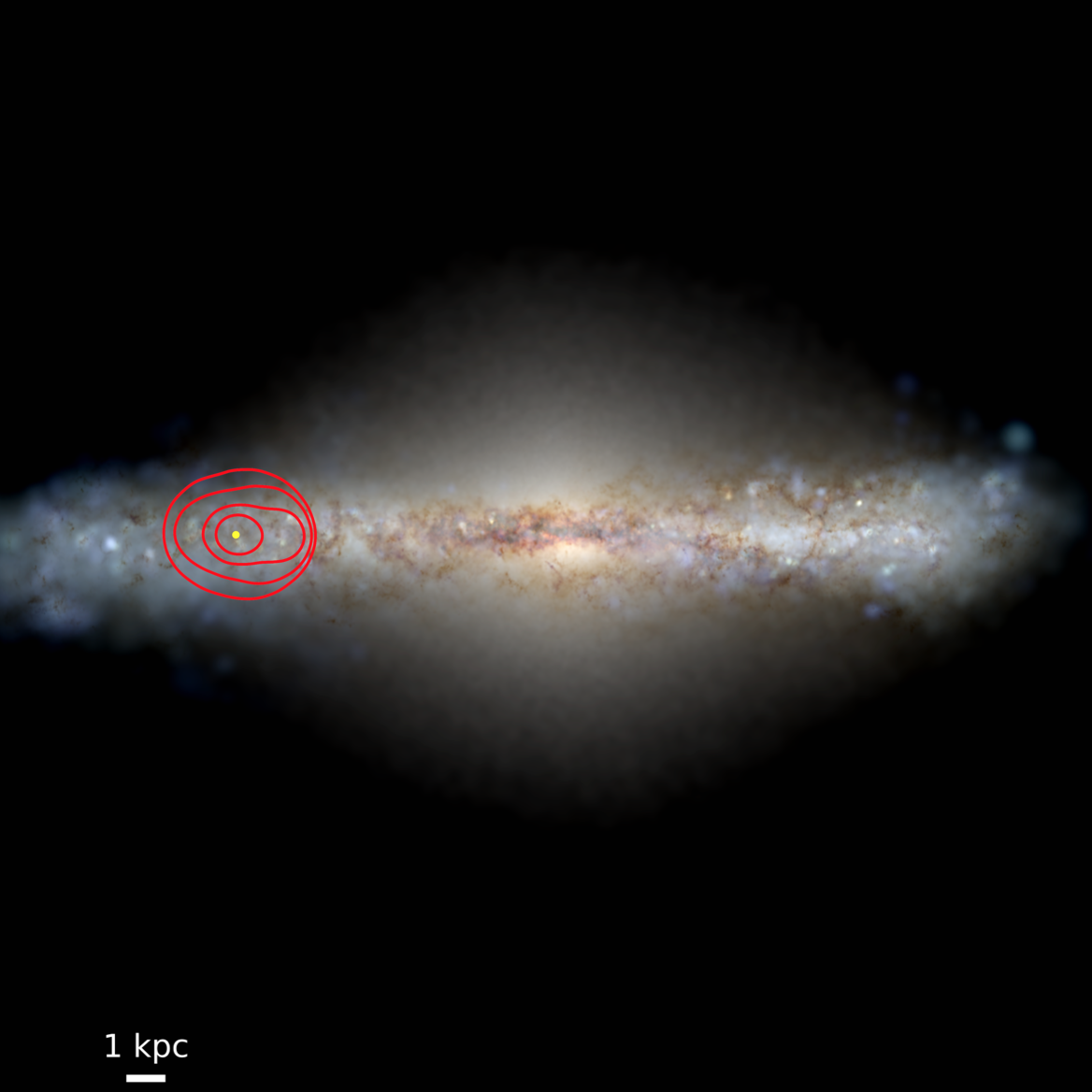}
\caption{Face-on (left) and edge-on (right) views of \mi, one of the simulated galaxies used to generate the mock catalog. The red contours represent density contours of the sample stars around the solar position that are used for constructing the Gaussian mixture model.}\label{fig:SN-m12i}
\end{center}
\end{figure*}

For each star the convolved abundances and velocities are drawn from a normal distribution around the ``true'' value generated in the synthetic stellar distribution, with the corresponding width, equivalent to the error on each property. Our mock catalog is thus essentially an all-sky version of the Gaia-APOGEE crossmatch. Although they are based on the synthetic Gaia surveys constructed in \citet{Sanderson19.firegaiadr2}, we refer to these simulated observations as \emph{mock catalogs} rather than \emph{synthetic surveys} since we do not overlay the APOGEE selection function. However, this will be done in future work to produce true synthetic surveys for the Gaia-APOGEE crossmatch. Finally, before running the mixture model we make the same quality selections on the mock catalogs as the observed catalog. These selections are all based on the simulated Gaia observations, not on our rudimentary APOGEE error modeling. Figure \ref{fig:SN-m12i} shows face-on and edge-on views of \mi simulation with density contours of the sample stars.

In addition to the APOGEE error model for elemental  abundances, we added the Two Micron All Sky Survey (2MASS) $\mathrm{J}, \mathrm{H}, \mathrm{K_s}$ magnitudes to our catalogs \citep{Skrutskie_2006}, corresponding to the bands used to select APOGEE targets \citep[see descriptions in][]{zasowski_2013,zasowski_2017}. The 2MASS photometric errors are estimated by using an exponential plus constant model and similar to the elemental  abundances, the convolved magnitudes are drawn from a zero-mean normal distribution with the corresponding variance. 

\section{Mixture models of mock Gaia-APOGEE catalogs}
\label{sec:mock}

To construct the mixture model, we use the three components of the space velocity in Cartesian coordinates $\{V_X, V_Y, V_Z\}$ and add iron abundances relative to solar, [Fe/H], as a fourth dimension. Compared to the classic Toomre diagram ($V_{XZ}$--$V_{Y}$ or $V_{RZ}$--$V_{\phi}$ plane), which has historically been used to separate different stellar kinematic components in the Solar neighborhood, this feature space allows for the possibility that stars in different kinematic components can have different metallicity distributions, but notably does not make any assumptions about what those distributions are. It likewise makes no assumptions about approximate symmetries in the phase space distribution (spherical, axisymmetric, or otherwise). This leaves us free to interpret the components obtained by the model in the context of broader ideas about galaxy formation, such as the expectation that stars in the thin disk, with a velocity distribution centered most closely on the Sun's, should also be the youngest and most metal-rich.

\begin{figure}
\includegraphics[width=\columnwidth]{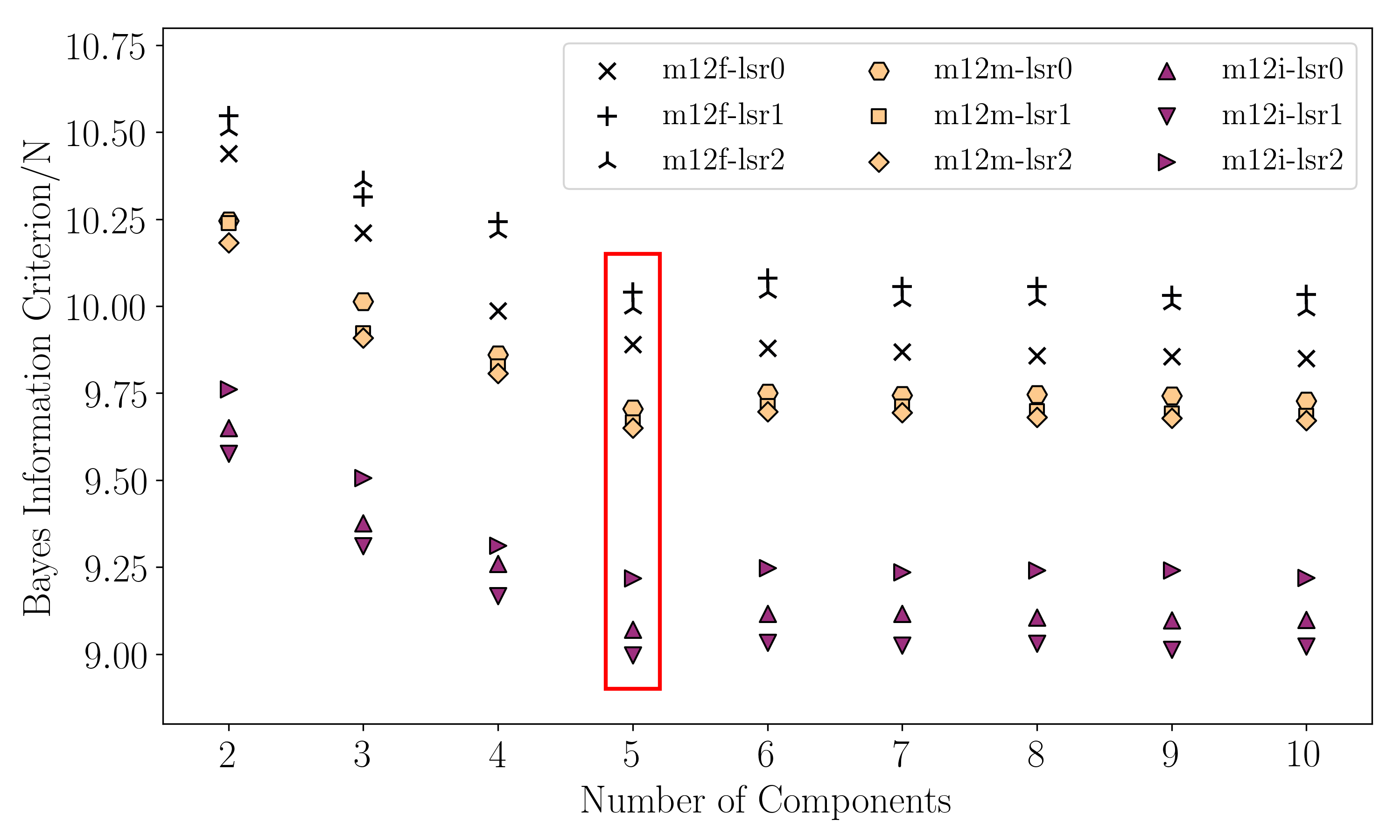}
\caption{Bayes information criterion (BIC) over the number of objects versus the number of components of mixture models for all nine Gaia-APOGEE mocks. Black symbols are for \mf simulation, orange symbols for \mi simulation, and magenta symbols are for \mm. All of the 3 simulations $\times$ 3 neighborhoods show the same BIC plateau at 5 components, which means the optimal decomposition features five components in all the simulations.}
\label{fig:mockBIC}
\end{figure}

Figure \ref{fig:mockBIC} shows the results of testing different numbers of Gaussian components to model this four-dimensional ``augmented Toomre diagram.'' There is a clear improvement up to 5 components and nearly no appreciable improvement after that (Elbow rule). Given the preference for a low number of distinct components (consistent with the idea of a thin disk, thick disk, and halo perhaps broken into some subpopulations) we choose the 5-component model for further examination.

\begin{figure}
\includegraphics[width=\columnwidth]{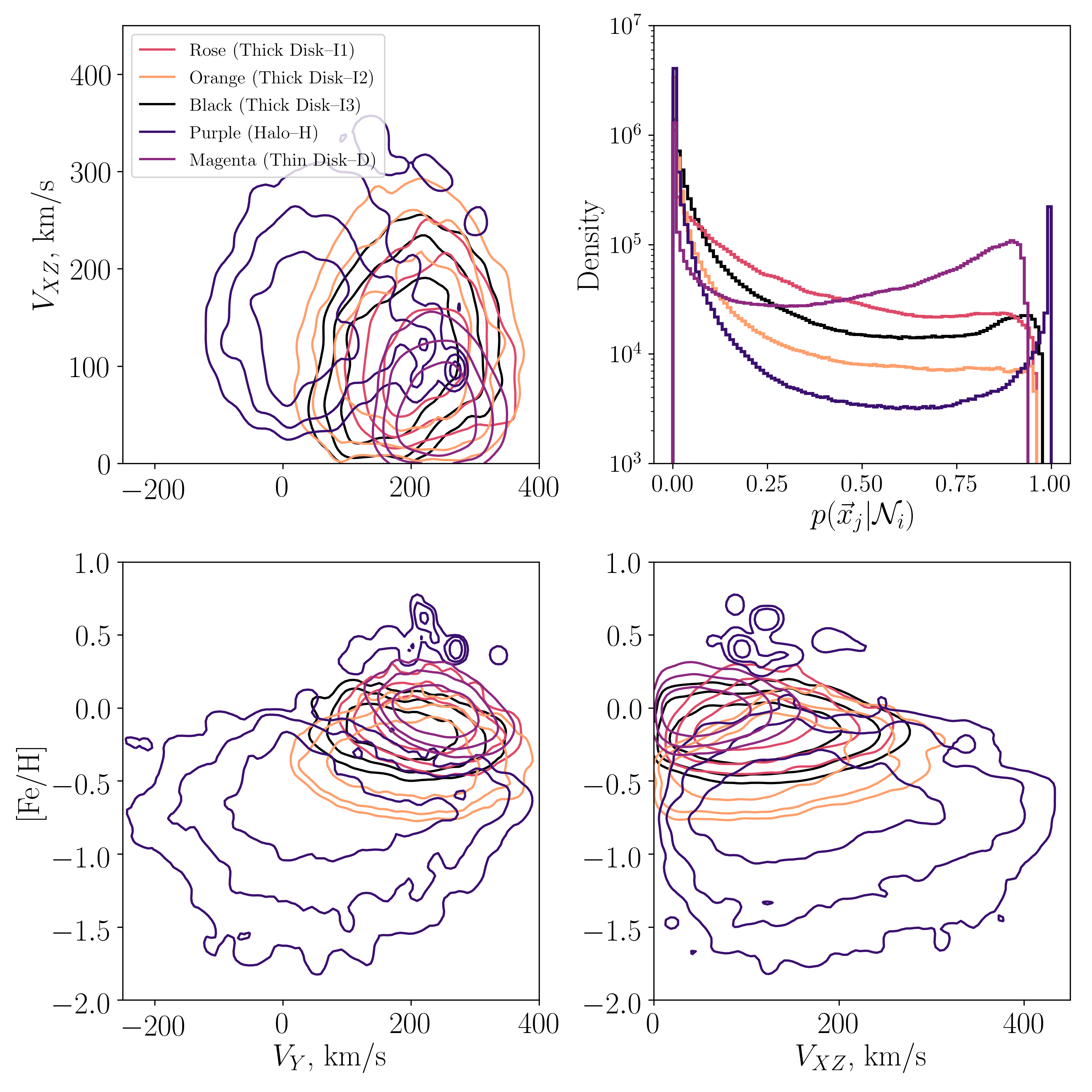}
\caption{Augmented Toomre diagram ($V_Y$ versus $V_{XZ}$ versus [Fe/H]) contours of the best-fit five-component mixture model to the \mf -LSR0 mock Gaia-APOGEE catalog. The top right panel shows the distribution of probabilities of belonging to each component for the different stars in the sample. In this model, two components can be identified unambiguously as the thin disk (magenta: metal-rich component with a narrow velocity distribution around the solar velocity) and another as the halo (purple: metal-poor component with a broad rotational velocity distribution around zero). Additionally, there are three progressively broader and more metal-poor components that together span the difference between these.}
\label{fig:mock5}
\end{figure}

In Figure \ref{fig:mock5} the density contours of each component of the best-fit model for one of the mock catalogs are shown in three projections of Toomre+[Fe/H] space. Each set of colored contours shows the density distribution of synthetic stars for which the probability of belonging to that component is highest. We see that this model includes components that fit the standard expectations of galaxy formation and the Solar neighborhood distribution: a metal-rich component with a narrow velocity distribution around the solar velocity (the ``thin disk,'' shown in magenta), a very metal-poor component with a broad rotational velocity distribution around zero (the ``halo,'' shown in dark purple), and a few progressively broader and more metal-poor components that together span the difference between these (shown in orange, black, and rose). We see a similar result, with slight variations in the relative positions of the components, for all nine mock catalogs.

The probability panel (top right) in this figure gives a sense of how well the model describes the data. It shows the distribution of probabilities of belonging to each component for the different stars in the sample. For some components, this distribution has well-defined peaks at 0 (i.e. the star is definitely not in that component) and 1 (i.e. the star is almost certainly part of that component) with a relatively low number of probabilities at intermediate values. This means that for these components, stars are easily classified. The halo and thin disk components demonstrate this behavior.

On the other hand, for other components, there is still a peak at zero but the distribution has more intermediate probability values and drops to zero before reaching $p=1$. This indicates that it is more difficult for the model to securely classify stars in one of these components; in this case it is because they are relatively similar as can be seen by examining the other panels, so many stars overlap with these. The three intermediate components demonstrate this behavior. This ambiguity between the components could come from the fact that a single Gaussian distribution in velocity is a bad description of the asymmetric velocity distribution displayed by thick disk stars, which is better modeled by a superposition of Gaussians (e.g. \citet{1907NWGot...5..614S,2004A&A...418..989N}).  In this case one would expect that multiple otherwise similar components are being used to effectively expand the velocity distribution in a basis of Gaussians. If the GMM is simply to be used to separate the three traditional constituents of the Solar neighborhood, the three intermediate components can be lumped together as the ``thick disk'' with no practical or conceptual difficulty. However, the preference for a small number of intermediate components, and the consistency of that number across all simulated and real datasets, also raises the intriguing possibility that the GMM is identifying subpopulations of stars with different intrinsic properties or different origins. We will evaluate this possibility, and analyze the intermediate components in detail, in \S\ref{subsec:thickdisk}.

We can verify the correspondence between the traditional components of the Solar neighborhood and these Gaussian components by examining the distributions of various properties not used to derive the model. For example, if the most metal-rich component truly corresponds to a traditional ``thin disk,'' then stars in this component should have a narrow distribution of heights above the disk plane, young ages, enhanced alpha abundances, and formation distances close to their present-day locations. 

\begin{figure*}
\includegraphics[width=\textwidth]{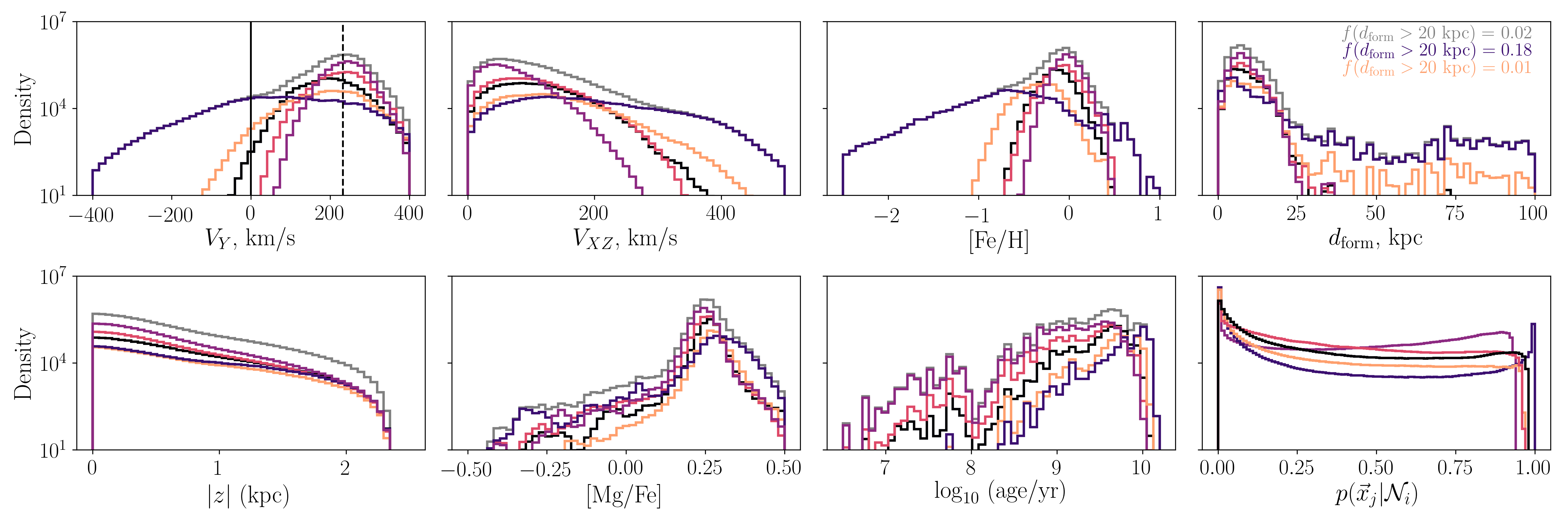}
\caption{Distribution of various properties of stars in each component of the best-fit mixture model, for the \mf-LSR0 mock catalog. As with all our mocks, one component (dark purple) is halo-like, one (magenta) is thin-disk-like, and three (rose, orange, and black) represent intermediate populations whose properties vary with LSR and formation history. The overall distribution for all stars in this sample is shown in grey. The upper right panel shows \dform distributions and stars with \dform $>$ 20 kpc are considered to be accreted. The accreted fraction of the total sample is about 2 percent and for the halo component it is about 20 percent. All five components have different age and alpha abundance distributions and these properties are not used to construct the model.}
\label{fig:mock5dists}
\end{figure*}

In Figure \ref{fig:mock5dists} we show a series of one-dimensional distributions of various stellar properties for each component in the same mock catalog. The probability panel is the same as the top-right panel of Figure \ref{fig:mock5} and colors of the components are the same as in that figure; the overall distribution is shown in grey where applicable. In contrast to Figure \ref{fig:mock5}, which uses the GMM as a classifier and thus includes only stars with $p_i(\vec{x}) > p_{j\neq i}(\vec{x})$ in each component, these distributions are calculated by weighting each star's contribution by its probability of belonging to that component. This illustrates the power of mixture modeling to permit full probabilistic analysis, which is especially important given the degree of overlap between populations (illustrated by the probability distributions shown in Figure \ref{fig:mock5}, which contain many intermediate values).

As seen in the top row of Figure \ref{fig:mock5dists}, the velocity distributions of these three intermediate components bridge the gap between stars rotating in the disk plane with the Sun ($V_Y=V_{Y,\odot}$, shown with a vertical dashed line) and stars orbiting in a broader distribution centered on the Galactic center. However, the inclusion of [Fe/H] as a fourth component shows that there is indeed some additional information in this distribution: components with lower mean metallicities which include metal-poor stars tend to have a broader velocity distribution. In the standard picture of galaxy formation, metal-poor stars originate predominantly from a galaxy merger and it is anticipated that they have a broader/more spheroidal velocity distribution. However, recent observations show that  the Milky Way's metal-poor stars have strong preference to be on prograde disk orbits \citep{Sestito2019, Sestito2020}. The similar behavior of metal-poor stars is shown for 11 of the 12 galaxies from the FIRE-2 simulation suite, including two of the simulations that we use in this work (\mf, and \mm) \citep{Santistevan2021}. The only galaxy among the 12 that does not show prograde orbits for metal-poor stars is \mi and the reason for this might be that all mergers occurred at about the same time, which could have distorted any coherent effect \citep{Santistevan2021}. This preference of metal-poor stars helps us to understand the velocity asymmetry that we see in Figure \ref{fig:mock5dists} for the halo component and the most metal-poor thick-disk-like component.

The other panels in Figure \ref{fig:mock5dists} show quantities that were not used in constructing the mixture model: 
\begin{itemize}
\item the formation distance \dform where the star particle from which each mock star was spawned relative to the main galaxy;
\item the height $|z|$ above the disk plane;
\item the magnesium-to-iron abundance ratio;
\item the stellar age.
\end{itemize}
These distributions can help us assess how the components identified by the model map onto ideas about the structure of the Solar neighborhood outlined in \S\ref{sec:intro}.

We track the approximate formation locations of each star particle in the simulated galaxy, relative to the center of the main halo at the time of formation, by post-processing the Gizmo\footnote{https://bitbucket.org/awetzel/gizmo\_analysis} snapshots saved from the simulation \citep{GizmoAnalysis}. We define \dform as the distance of the star particle from the host galaxy center in the first snapshot after it is formed. As in \citet{bonaca17} and \citet{sanderson17}, we consider stars with \dform $>$  20--30 kpc to be accreted depending on the simulation (see Figure 1 of \citet{sanderson17}), although the caveats discussed extensively in both those works apply here as well. In short, this can be considered a conservative definition of the accreted stellar component. 


Overall, we see that the distribution of the magenta component closely resembles what would be considered the thin disk: exclusively young, metal-rich stars formed inside the galaxy, with alpha-to-iron ratios close to solar, orbiting with the Sun near the disk plane. Likewise the dark purple component pretty clearly fits our expectations for the halo: velocities consistent with a kinematically hot spheroid (slightly counter-rotating in some cases), a broad spatial distribution, and exclusively old, metal-poor, alpha-enhanced stars. These two components are clearly identifiable in the best-fit mixture model for all nine mock catalogs. Given the clear parallels between these two components and standard interpretations of stellar populations, we will refer to them as the thin disk and halo for short in the remainder of the paper. For these two components the formation distance of the stars also supports classical theories about their origin: stars in the thin disk component all formed within 25 kpc of the Galactic center, while about 20 percent of the halo component was formed beyond 20 kpc, where for this simulation most material can be considered accreted rather than formed in situ. Interestingly, more than half of the stars in the halo component have an origin consistent with our picture of an early spheroid: extremely old ages and low metallicity, yet formed within the main galaxy. These stars come from early, extremely bursty epochs of star formation seen in the simulations \citep{2018MNRAS.480..652E, 2020MNRAS.494.1539Y, Muratov2015, Sparre2017, FaucherGiguere2018}. The accreted fraction the halo component of \mf, \mi, and \mm \- simulations is about 20\%, 10\%, and 25\% respectively, these percentages depend on assembly history of each simulation and these three simulations have different accretion histories. In each simulation, the accreted fraction of each component is almost the same (within 1--2 percent) in the three different LSRs and also the the accreted fraction of all stars is the same in all nine mocks (about 2 percent).

From the velocity and metallicity distributions, it is understandable how these two components, comprising the oldest and the youngest stars in the sample, are most easily picked out by the model with certainty. In the age distribution, which is not used in our model, they are also very well separated. They are most consistent with a Gaussian velocity distribution, though for two very different reasons: one has barely been transformed from its birth distribution at all while the other has evolved for many dynamical times. They also pick out the metal-poor tail and metal-rich peak of the total stellar distribution, respectively, making these stars particularly easy to classify.

\subsection{Thick Disk Decomposition}
\label{subsec:thickdisk}

The remaining three components, which together make up what would traditionally be referred to as the ``thick disk,'' lie intermediate to the thin disk and halo components in all the characteristics we examined. The separation of this intermediate population into multiple components could indicate less consistency with the assumptions of the model, a more complex origin, or some of both. In this section we discuss these three components in detail.

The component shown in orange most resembles the halo component in terms of its velocity, metallicity, and $z$ distributions, but differs in a few important respects. First, there are some younger stars present than in the halo component (purple); and second, although there are a few stars present formed at larger distances, the overall makeup of this sample resembles the total Solar neighborhood in terms of its accreted fraction (about 1 percent). The stars in this component are also slightly more metal-rich than the halo, and notably less alpha-enhanced.

\begin{figure}
\includegraphics[width=\columnwidth]{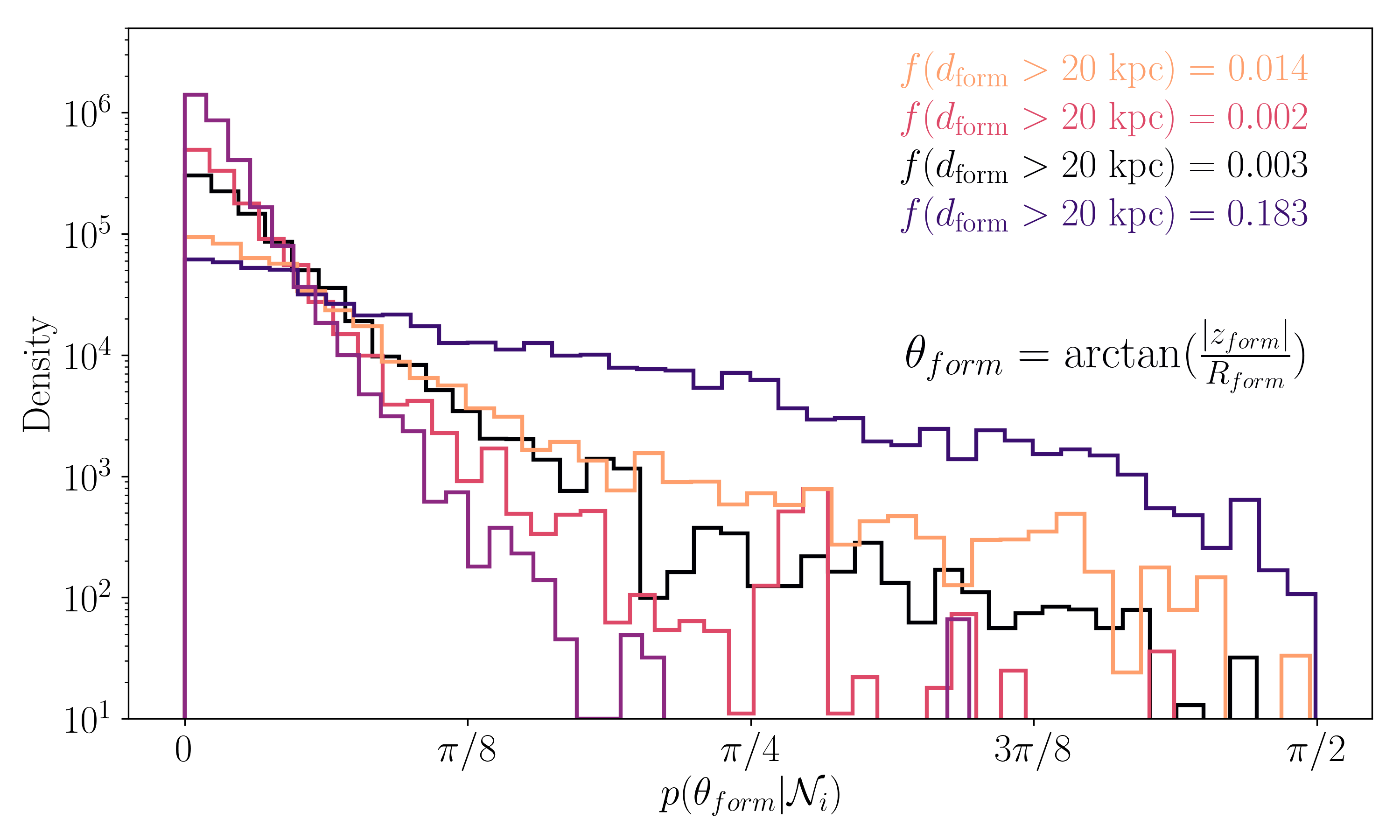}
\caption{The distribution of the inclination angle of stars relative to the disk plane at the time of formation for all components of the \mf-LSR0 mock catalog. The orange component has a much flatter distribution than the other two intermediate components (rose and black), but it is still not as flat as the distribution of the halo component (purple).} 
\label{fig:theta}
\end{figure}

Moreover, from Figure \ref{fig:theta}, which shows the distribution of the inclination angle of stars relative to the disk plane at the time of formation, we can see that this component has a much flatter distribution than the other two intermediate components, but it is still not as flat as the distribution of the halo component. The rose and black components also have distinct distributions in this view: most of the stars in the rose component have \thetaform$<\pi/4$ (i.e. are formed quite close to the galactic plane), while the black component has a longer tail at higher \thetaform.

\begin{figure}
\includegraphics[width=\columnwidth]{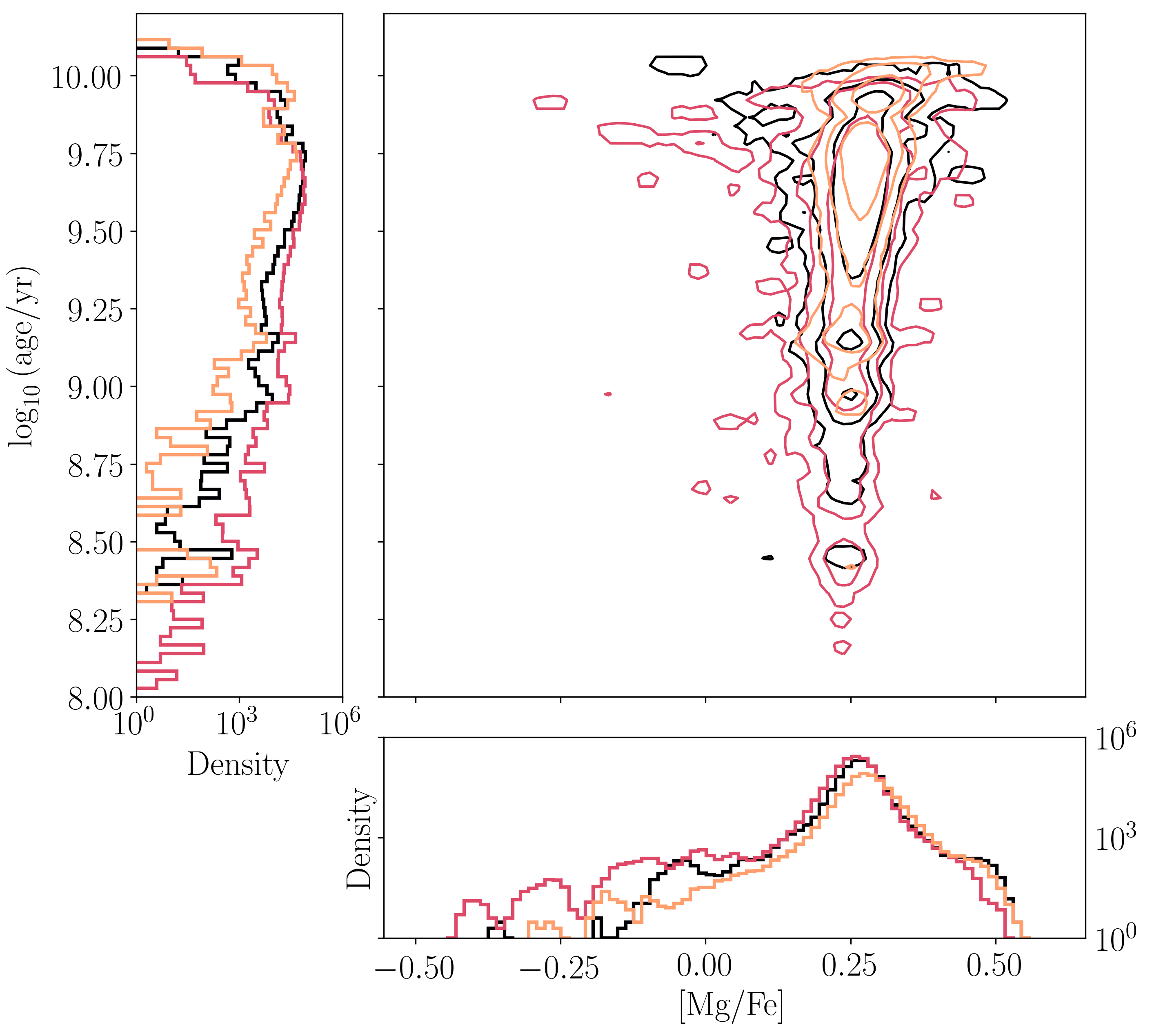}
\caption{The distribution of age-alpha for three intermediate (thick disk) components of the \mf-LSR0 mock catalog. The rose component has younger, less alpha-enhanced stars, which is consistent with the picture given by its formation angle distribution (Figure \ref{fig:theta}).}
\label{fig:age-alpha}
\end{figure}

The three components also show differences in their age and alpha abundance, two other features that were not used in classifying them. Figure \ref{fig:age-alpha} shows the distribution of age versus alpha for these three components, obtained by classified stars as shown in Figure \ref{fig:mock5}. The rose component has younger, less alpha-enhanced stars, which is consistent with the picture given by its formation angle distribution that these stars formed after the disk plane is well established and the disk relatively cold. In these simulations, this is usually due to an influx of cold, high-angular-momentum gas at relatively late times \citep{2017arXiv171203966G}. Conversely, the orange component includes the oldest and most alpha-enhanced stars of the three, consistent with the picture that these stars formed earlier when the disk was kinematically hotter (often because of a wider distribution of angular momentum in the cold gas accretion).

The GMM used to classify the stars in these components is based solely on 3D velocities plus iron abundance, yet we see that they have different age and [Mg/Fe] distributions and are formed in different locations relative to the galactic center (\dform) and to the disk plane at the time of formation (\thetaform). Furthermore, while the thin disk and halo components are consistently identified in every mock catalog, the characteristics of intermediate components identified in different simulations (which have different assembly histories) are markedly different. If the GMM were simply decomposing the asymmetric drift in a combination of Gaussians, it is highly unlikely that all these differences would be apparent. We thus argue that the mixture model is indeed identifying components with different physical origins within the thick disk.

\subsection{Dependence on position in the galaxy}

With mock catalogs, it is possible to study how the distribution of stars in the augmented Toomre space varies as a function of position in the Galaxy. The solar position within the simulation is a fairly arbitrary choice since the simulated galaxy does not resemble the Milky Way in its detailed structure, such as the position and number of spiral arms, the size and orientation of a bar, or the number and location of tidal streams in the halo. To illustrate how the augmented Toomre diagram changes as a function of location, and to test the sensitivity of this approach to the choice of solar position, we generated different catalogs for three solar locations in each simulated galaxy. They have 120 degree displacement with each other on the solar circle. As shown in Figure \ref{fig:mockBIC}, the behavior of the BIC and the number of preferred mixture components is consistent across different solar locations; implying that this technique will be successful when applied to the real data regardless of the local variations in the density or distribution of stars.

Figure \ref{fig:dform-age} shows that the characteristics of the components identified by the model are not completely the same in all of the nine mock catalogs, but in all of them we have three components for the thick disk. In addition to position dependence in the galaxy, these differences can depend on assembly history of each simulation since these three galaxies have different relative formation times and accretion histories. 

\begin{figure*}[t]
\centering
{\large m12f}\\
\includegraphics[width=\textwidth]{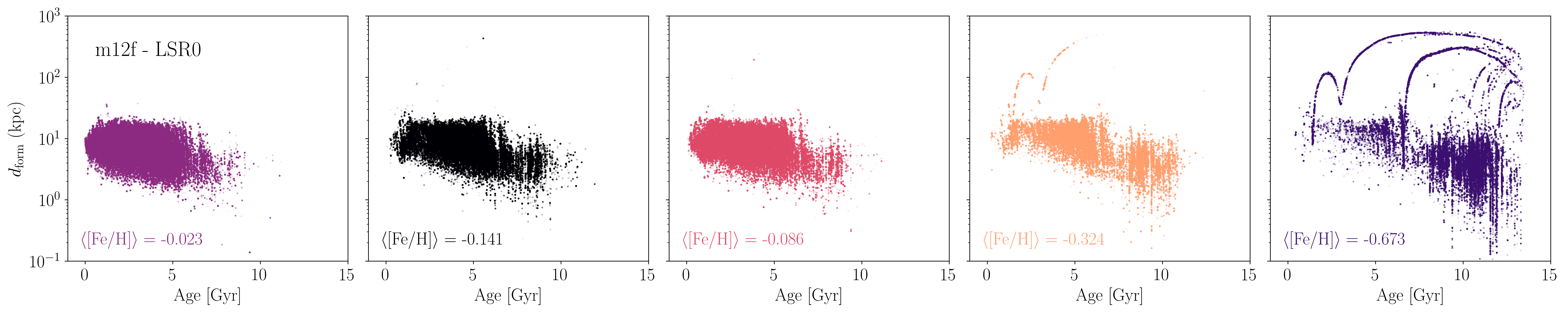}\\
\includegraphics[width=\textwidth]{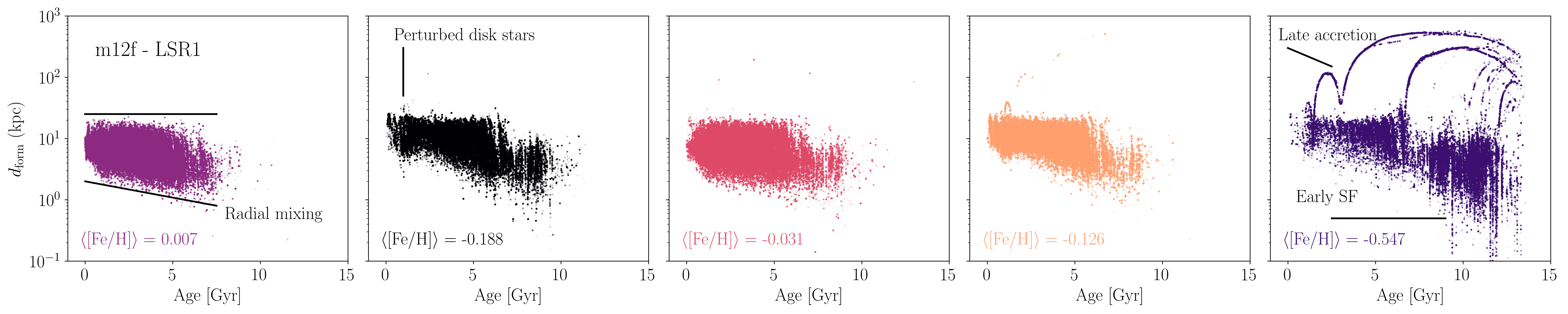}\\
\includegraphics[width=\textwidth]{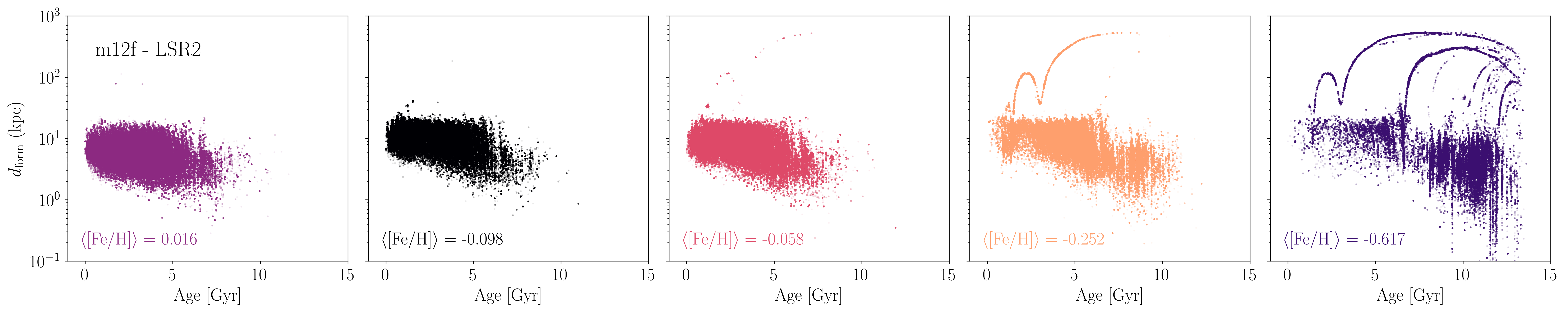}\\
{\large m12i}\\
\includegraphics[width=\textwidth]{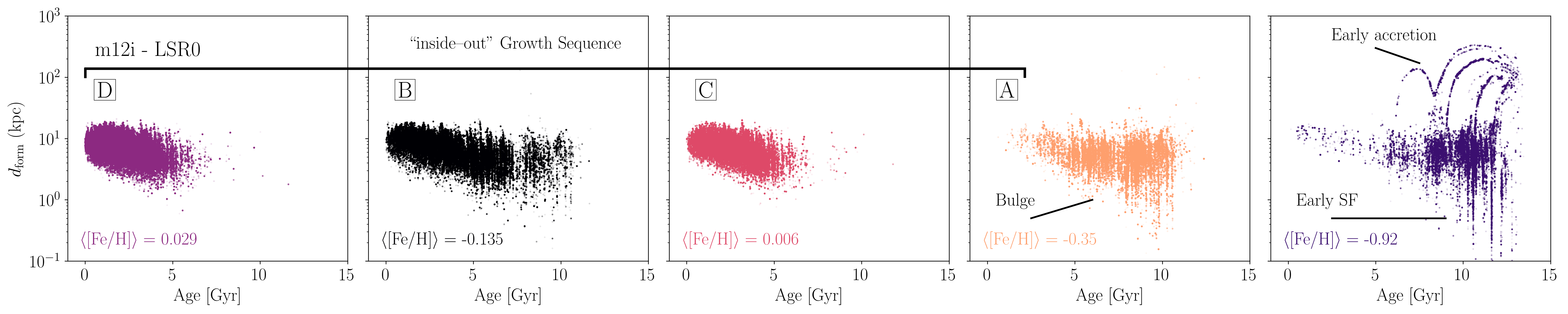}\\
\includegraphics[width=\textwidth]{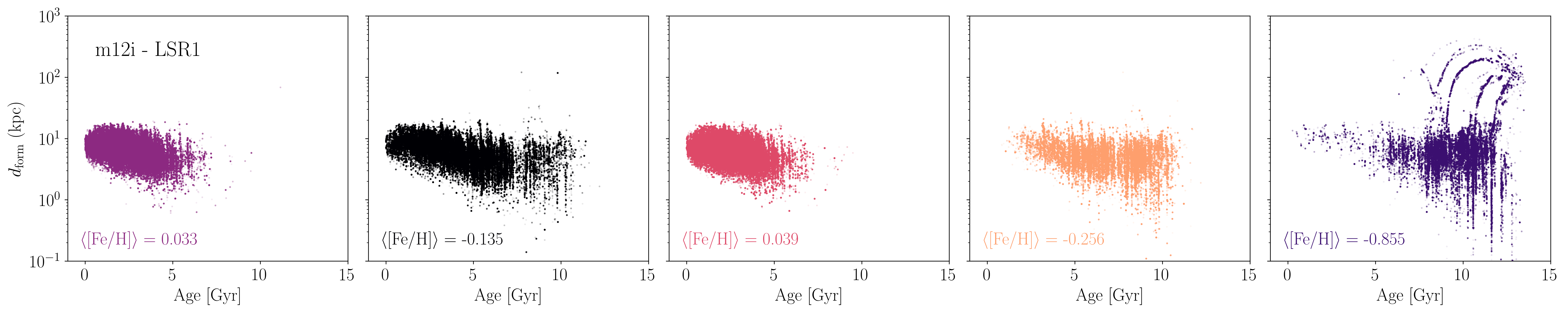}\\
\includegraphics[width=\textwidth]{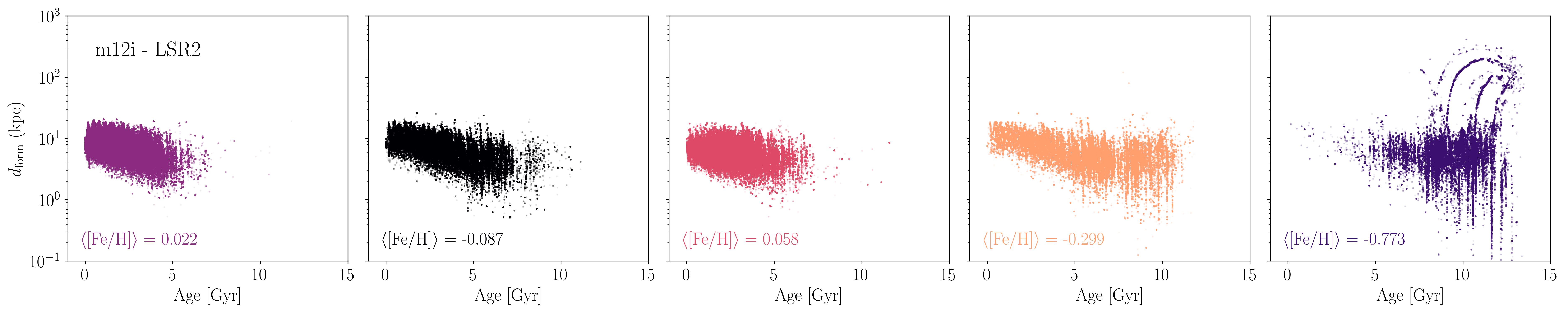}\\
\end{figure*}

\begin{figure*}[t]
\centering
{\large m12m}\\
\includegraphics[width=\textwidth]{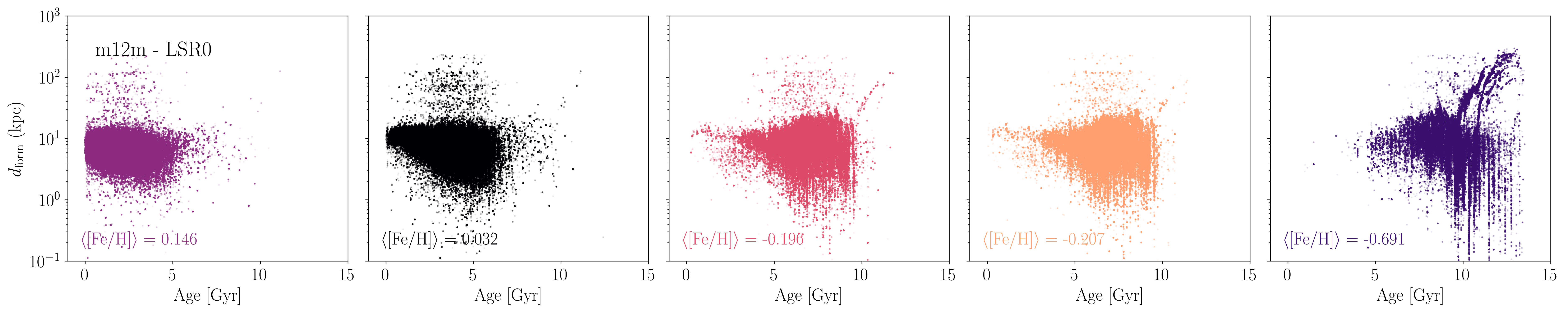}\\
\includegraphics[width=\textwidth]{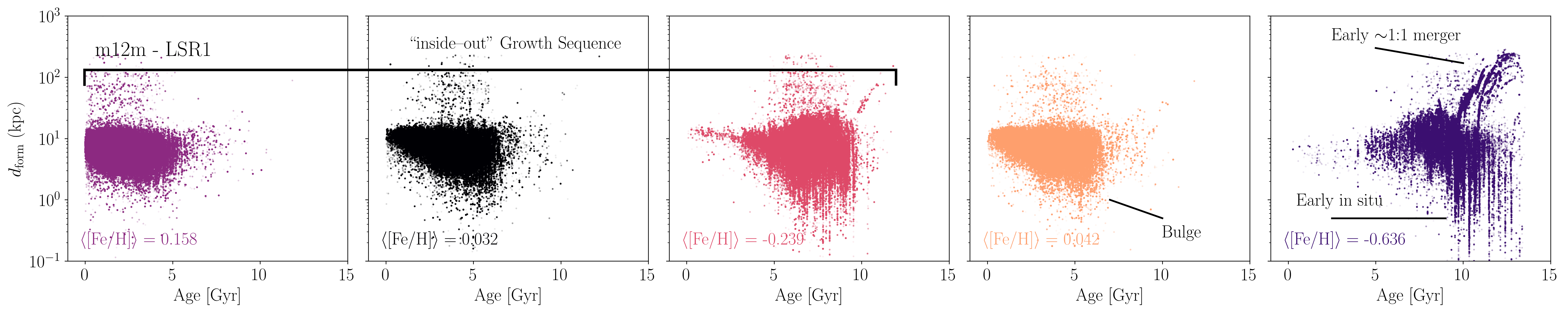}\\
\includegraphics[width=\textwidth]{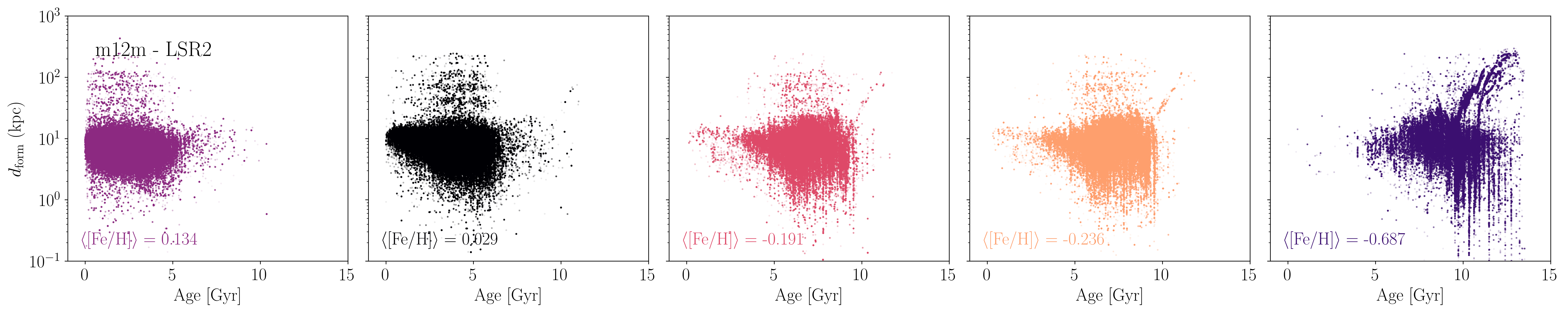}\\
\caption{Formation distance - age for each component in all simulations. Some major events in each simulation are annotated in each panel (e.g. radial mixing, ``inside--out'' growth sequence, merger, etc.) The intermediate components show the stages of inside-out formation in the disk, modified by merger interactions that scatter young stars formed near the solar circle, by the rotation of the disk during the merger, and by ``blurring'' (the selection effect that Solar neighborhood stars are preferentially near apocenter). In \mf simulation the intermediate components contain stars that have been heated by interactions with satellite galaxies such as Sagittarius or a late merger. The halo components of \mm panels support the idea that the majority of the halo component is from one or two early, massive mergers.}
\label{fig:dform-age}
\end{figure*}




\subsection{Dependence on assembly history}

Many scenarios have been proposed for the origin of the thick disk. It may emerge from stars migrating outward from the hot, inner disk \citep[e.g.][]{Loebman2011}, from a turbulent interstellar medium (ISM) \citep[e.g.][]{Bournaud2009} or a gas-rich merger \citep[e.g.][]{Brook2004} at high redshift, from a satellite dynamically heating a preexisting stellar disk \citep[e.g.][]{Villalobos2008}, or from the accretion of stars stripped from satellites \citep[e.g.][]{Abadi2003}. In addition to each of these single-origin theories, the thick disk could also arise from various combinations of these processes at different times. Moreover, in these simulations, since there was not always a single main progenitor, the definition of in/ex-situ for for early-forming stars is complicated \citep{2020MNRAS.497..747S}. 

Figure \ref{fig:dform-age} shows formation distance of stars versus their age in each component. For each simulation we can see the dependence on position in the galaxy (different LSRs) and by comparing simulations to each other, we are able to see the dependence on assembly history. In addition, the movies available here\footnote{https://web.sas.upenn.edu/dynamics/data/ananke-2a} show the formation and spatial evolution of each component over time.

\begin{figure}
\includegraphics[width=\columnwidth]{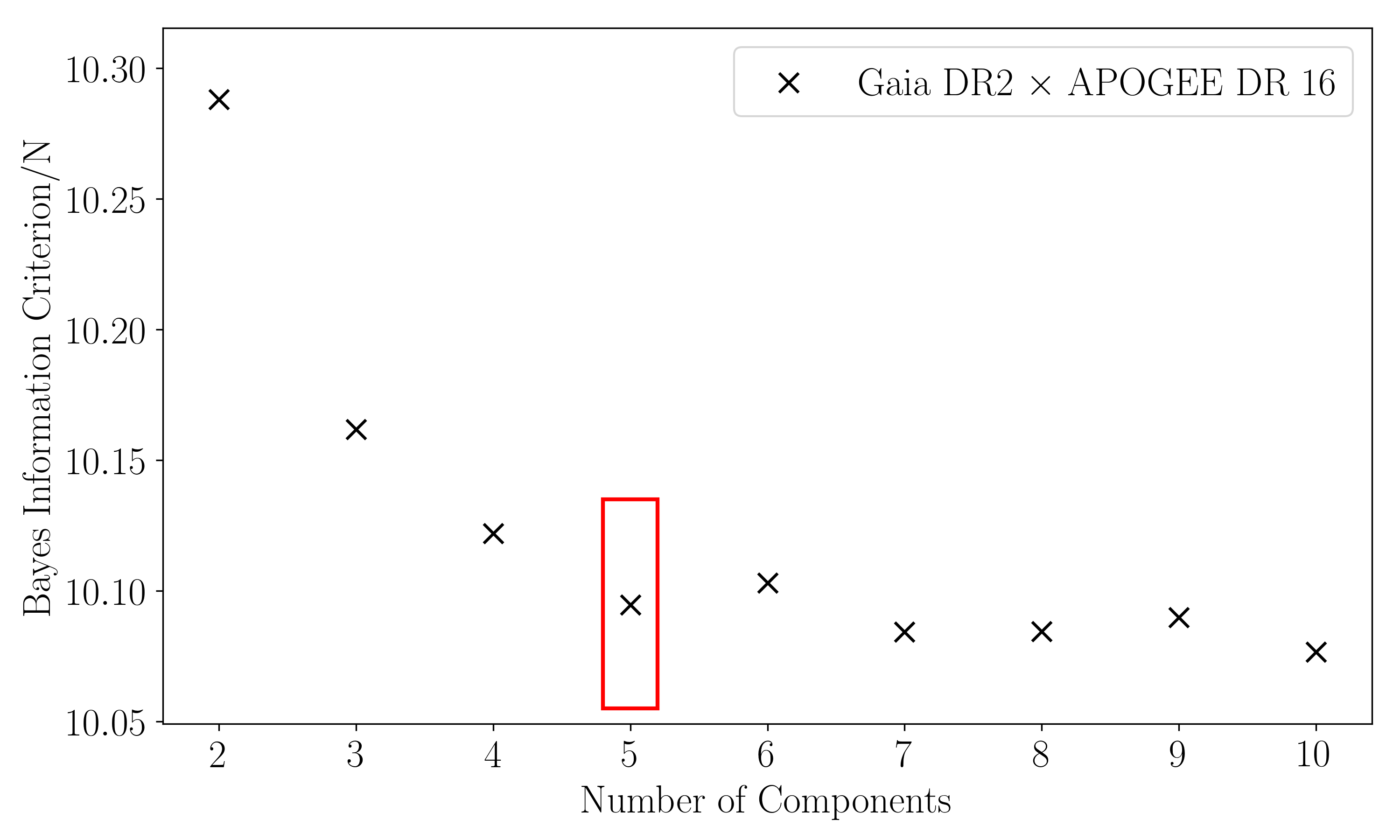}
\caption{Bayes information criterion (BIC) over the number of stars versus the number of components of mixture models for the real Gaia DR2 - APOGEE DR16 crossmatched catalog (compare to Figure \ref{fig:mockBIC} for the mock catalogs).}
\label{fig:realBIC}
\end{figure}

In \mf\ (top 3 rows), we see that the halo component is clearly accreted, but includes the very earliest star formation in the main halo as well. The oldest thick disk component (orange) is mostly composed of stars that formed very early on near the galaxy center and radially migrated outward \citep{2016ApJ...820..131E, 2017MNRAS.467.2430M, 2018MNRAS.480..652E}. It also has some stars from a merger that comes in on a nearly co-planar orbit late in the simulation (bouncing track at high \dform). This leads to a far lower mean metallicity for this component than for the other two thick disk components (black and rose). These also mostly form interior to the solar circle, but at more intermediate radii, starting and finishing their star formation later than the orange component. The movie of these components shows they also have some initial diskiness. The disk in this simulation starts out perpendicular to what ends up being the disk plane at present day, and is torqued by a merger into its present configuration. These intermediate components thus show the stages of inside-out formation in the disk, modified by merger interactions that scatter young stars formed near the solar circle, by the rotation of the disk during the merger, and by ``blurring'' (the selection effect that Solar neighborhood stars are preferentially near apocenter). These two intermediate components are also staggered in age, especially in LSR0, and this is reflected in the systematic variation of their mean [Fe/H]. Closer to $z=0$, we see that the two later-forming thick disk components also contain stars that were scattered onto orbits that intersect the solar circle by the late interaction with the merging galaxy, especially in LSR1 and LSR2. The thin disk forms latest, after the merger torques all the thick disk stars vary rapidly over into its preferred plane from a nearly 90-degree angle. There is blurring here as well in that the stars in the thin disk mostly come from interior to the solar circle, but the central part is almost empty since these are stars on the most circular orbits. This component is also scattered by the merger; it looks like perhaps some stars are removed due to the interaction, transferred to the higher-dispersion component.

\begin{figure*}
\includegraphics[width=\textwidth]{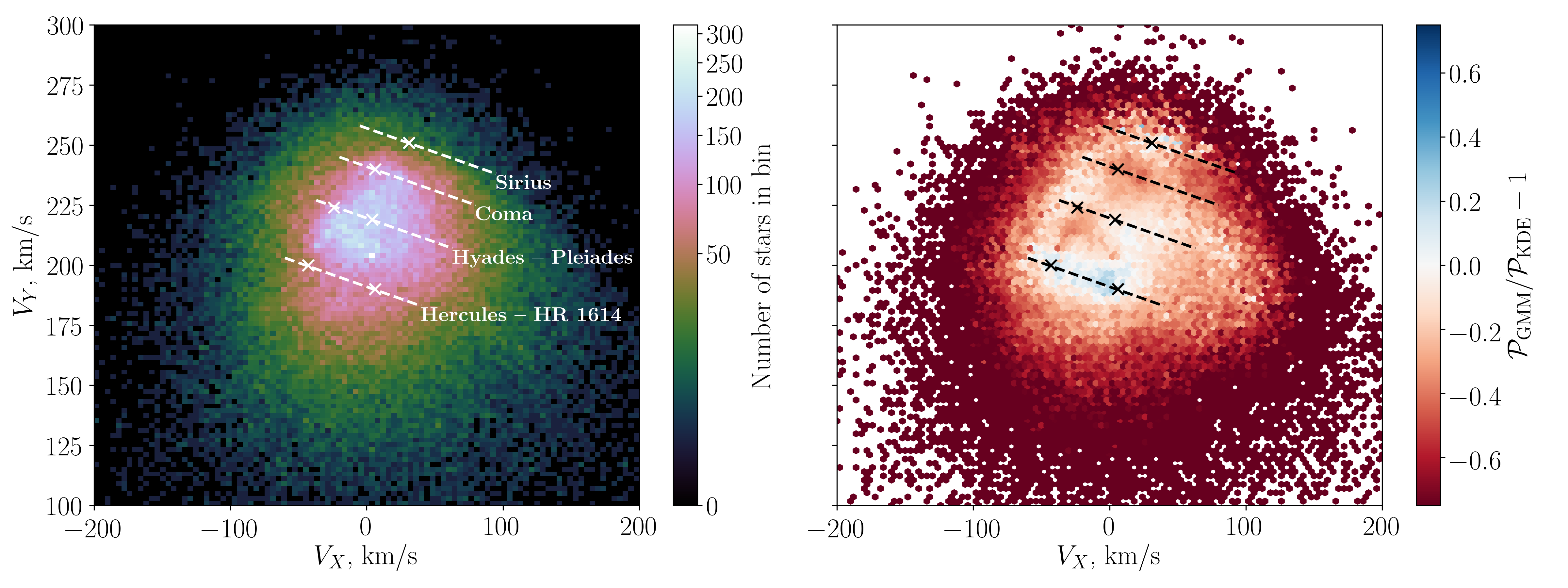}
\caption{Left panel: 2D histogram of the velocity plane for the stars in the solar neighbourhood. Right panel: residual plot for the GMM model in the velocity plane. The positions of the centres of the stellar moving groups are shown in the residual plot. The dashed lines show the approximate trace of the branches and they are at the same levels in both plots. They show that over[under]-estimation of the density occurs at the locations of physical under[over]densities.}
\label{fig:residual}
\end{figure*}

Examining \mi (middle three rows), we see that its different history is reflected in the classified components, but that the same set of mechanisms is present: accretion and early star formation in the halo component, young stars and blurring in the thin disk, and radial migration and heating in the three thick disk components, which pick out stars of different ages formed in different regions: the most metal-poor component (orange) is also the oldest and most transformed by radial migration; the intermediate component (black) shows the classic combination of blurring and radial transport; the youngest and most metal-rich (rose) actually resembles the thin disk in some respects, but its stars are mostly older, have a broader velocity distribution, and show a greater alpha enhancement. Here the different solar positions do not differ as much as in \mf, a reflection of this simulated galaxy's overall much calmer late-time history that leads to a well-mixed system. This is also reflected in the larger difference between the halo metallicity and the disk-like components, and in the systematic variation of the mean metallicities of the disk-like components. 

\begin{figure}
\includegraphics[width=\columnwidth]{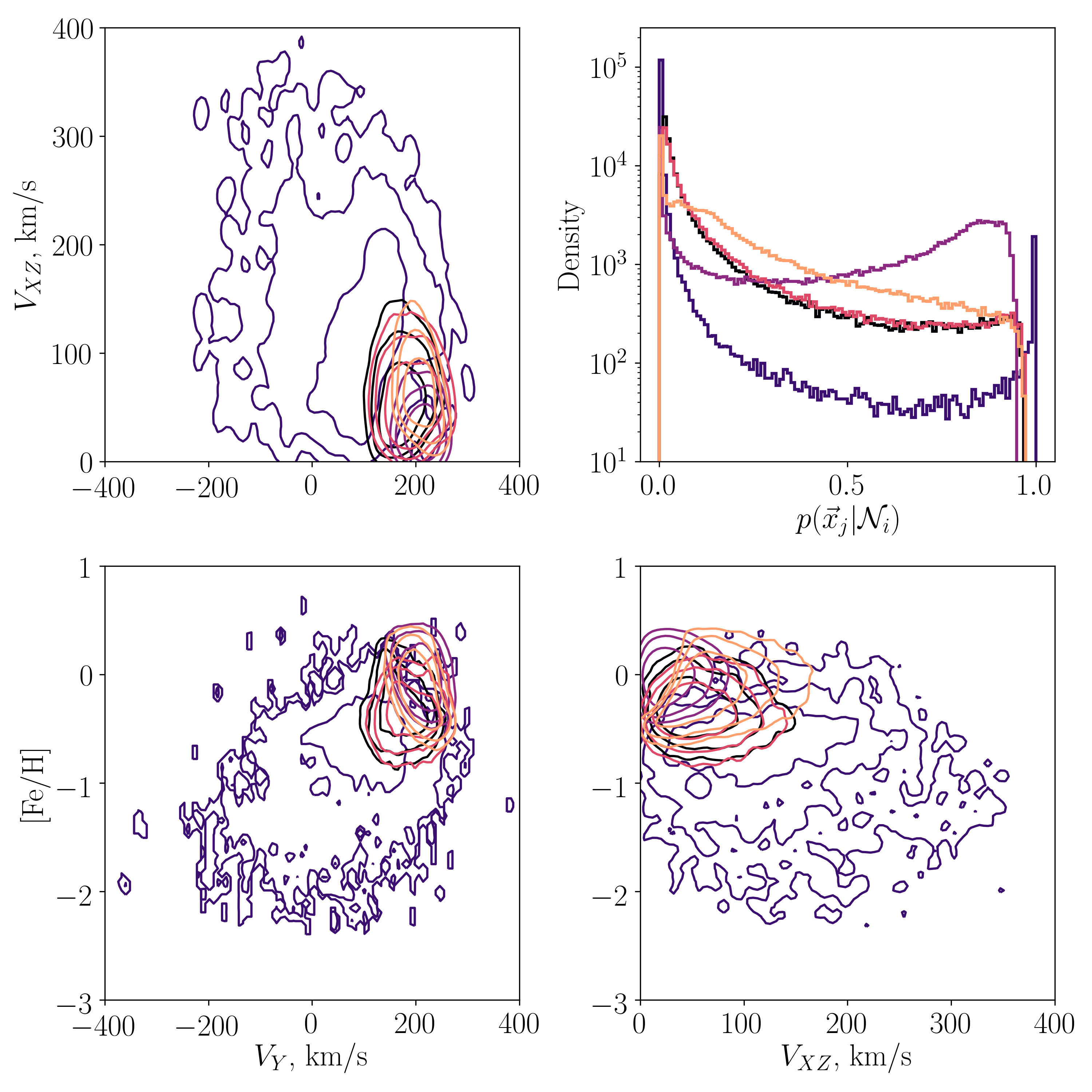}
\caption{Augmented Toomre diagram for the 5-component mixture model of the real Gaia-APOGEE catalog (compare to Figure \ref{fig:mock5} for the mock catalog).}
\label{fig:real5}
\end{figure}

Finally, in \mm\ we find yet another superposition of the different formation channels. Here we see stars with high formation distance even in the thin disk component, and across all the others. These stars come from a $\sim$1:1 merger at $z \sim 2$ that results in a starburst across the whole galaxy; a handful of these even end up on thin-disk-like orbits.\footnote{Since the merger is roughly equal mass, it is somewhat arbitrary which galaxy is the main progenitor.} There are traces of this merger scattered among all the components but the bulk of its stars are in the halo (dark purple) where it is apparent as a thick descending line at high age. We also see some variations in the makeup of the thick disk components at different solar positions: in two cases (LSR0 and LSR2) there are two components with similar average metallicities of around $-0.2$ that are very old stars likely related to the merger starburst, and another population at much higher metallicity (around 0.0) that looks more like a classic radially migrated distribution. In the other case (LSR1) the proportions are reversed, and so are the average metallicities (now two components have [Fe/H]$\sim$0 and one has [Fe/H]$\sim-0.2$). 

To summarize, we find that across all simulations and solar locations, the components of our mixture models generally correspond to different formation channels for their stars, including the decomposition of the thick disk into multiple subcomponents. These can be disambiguated to some extent by the grouping of the mean metallicities of the different components, which varies based on the specific formation channels involved for each galaxy, or by examining distributions of other elemental  abundances if these are available.

\section{Mixture models of the real Gaia-APOGEE catalog}
\label{sec:real}

We next constructed a mixture model of the \emph{real} Gaia DR2-APOGEE DR16 crossmatched catalog. As before, we tried models with different numbers of Gaussian components and used the Bayes Information Criterion to choose a preferred number of components. Figure \ref{fig:realBIC} shows that as with the mock catalogs, the model with 5 components provides a significant improvement over fewer-component models, while adding additional components improves the performance far less. We therefore proceed with 5 components as for the mock catalogs.

\begin{table*}
\begin{center}
\begin{tabular}{llccc}
\hline
\hline
Component & Color & $\tau$ & $\vec{\mu}$ [km/s, km/s, km/s, dex] & $\hat{\mathbf{\Sigma}}$ \\
\hline
H & Dark Purple & 0.04 & 
$\left[ 19.01, 99.30 ,  3.32, -0.69 \right]$ &
$\left(\begin{array}{rrrr}
13107.177 & -447.250 & -851.968 &  4.756 \\
  -447.250 & 7498.934 &   -2.700 & 15.538 \\
  -851.968 &   -2.700 & 4212.157 &  1.404 \\
     4.756 &   15.538 &    1.404 &  0.172
\end{array} \right)$ \\
I1 & Rose & 0.14  & 
$\left[ -8.16, 195.53,  18.5 ,  -0.28 \right]$ & 
$\left(\begin{array}{rrrr}
2667.817 &  -81.931 & 132.633 & -1.146 \\
  -81.931 & 1165.986 & -98.953 &  0.355 \\
  132.633 &  -98.953 & 648.367 & -1.070 \\
   -1.146 &    0.355 &  -1.070 &  0.045
\end{array} \right)$ \\

I2 & Orange & 0.22 & 
$\left[ 44.25,  209.84, -2.24,  -0.10 \right]$ & 
$\left(\begin{array}{rrrr}
1857.374 & -405.721 &  -2.581 &  1.080 \\
 -405.721 &  540.602 &   4.405 & -1.355 \\
   -2.581 &    4.405 & 423.121 &  0.009 \\
    1.080 &   -1.355 &   0.009 &  0.041
\end{array} \right)$ \\

I3 & Black & 0.13 & 
$\left[ -2.16, 182.16, -19.73,  -0.25 \right]$ & 
$\left(\begin{array}{rrrr}
3368.552 &  -60.419 & -157.776 & -1.134 \\
  -60.419 & 1164.563 &  -35.883 &  0.532 \\
 -157.776 &  -35.883 &  711.881 &  1.784 \\
   -1.134 &    0.532 &    1.784 &  0.059 
\end{array} \right)$ \\

D & Magenta & 0.46 &
$\left[ -1.28,  211.78,  -0.37,  -0.01 \right]$ & 
$\left(\begin{array}{rrrr}
1019.530 & 142.882 &   1.035 &  0.200 \\
  142.882 & 441.262 &  14.896 & -1.027 \\
    1.035 &  14.896 & 201.201 & -0.012 \\
    0.200 &  -1.027 &  -0.012 &  0.026
\end{array} \right)$ \\

\hline
\hline
\end{tabular}
\end{center}
\caption{Best-fit Gaussian mixture model for the real Gaia-APOGEE catalog.}
\label{tbl:realModel}
\end{table*}

In Table \ref{tbl:realModel}, we present the coefficients of the five-component Gaussian mixture model trained on the real Gaia-APOGEE dataset (different components of the model are defined by their colors). This trained model can be used to identify members of these components in other surveys (\S \ref{sec:discussion}). The correlation matrices in this table show that for the thin disk component there is a strong correlation (large off-diagonal elements) between $V_X$ and $V_Y$, pointing towards a cylindrical symmetry and $V_R$, $V_\phi$ as coordinates. However, for other components this is not as strong; the halo component in particular is correlated similarly strongly in $V_X$ and $V_Y$ and in $V_X$ and $V_Z$, which suggests spherical symmetry as expected.

\begin{figure*}
\includegraphics[width=\textwidth]{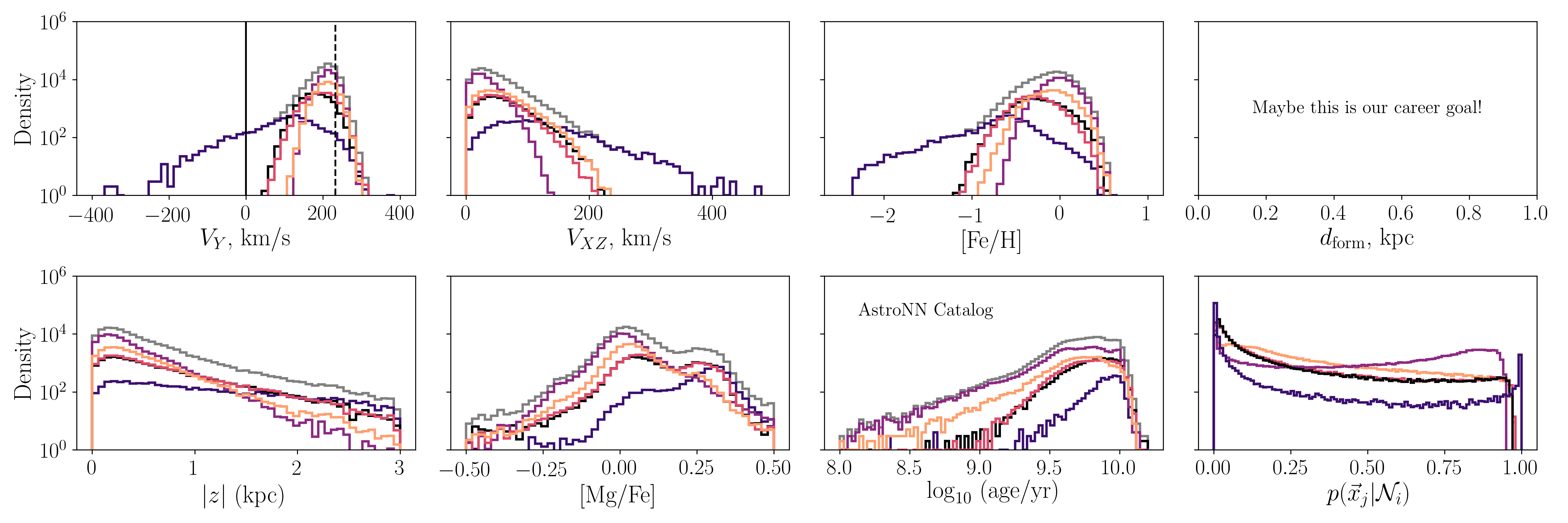}
\caption{Distribution of various properties of stars in each component of the best-fit mixture model, for the real Gaia-APOGEE catalog. We do not have the \dform values and ages for the real catalog, APOGEE provides magnesium-to-iron ratios, Gaia provides heights above the disk plane at present day, and the astroNN Value-Added Catalog provides ages.}
\label{fig:real5dists}
\end{figure*}

\begin{figure}
\includegraphics[width=\columnwidth]{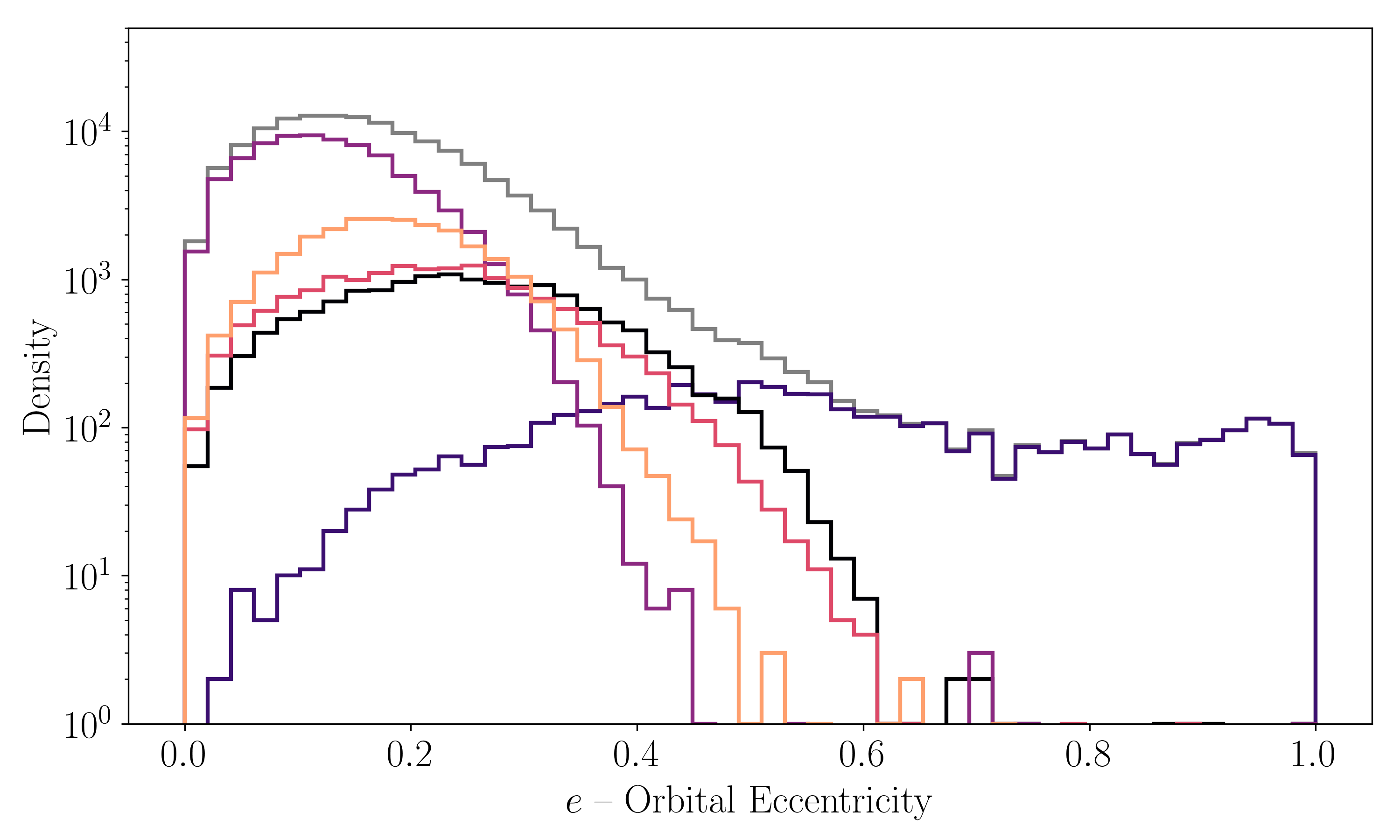}
\caption{The distribution of orbital eccentricity from astroNN catalog. Stars of the halo component have more elliptical orbits and thin disk stars are on circular orbits. The three thick disk components in this space are also different from each other.}
\label{fig:astroNN-eccen}
\end{figure}

In Figure \ref{fig:residual}, the left panel shows the residual plot of our GMM model in the $\{V_X, V_Y\}$ plane, and the right panel is the density plot of the stars in the solar neighbourhood. In the residual plot, we compare the kernel density estimation (KDE) of the stars in that plane with the GMM probability estimation, which shows the over- and under-estimation locations of the density. The positions of the centres of the stellar moving groups according to \cite{Antoja2008, Antoja2010} are shown in the residual plot. The dashed lines show the approximate trace of the branches and they are at the same levels in both plots. They show that over[under]-estimation of the density occurs at the locations of physical under[over]densities. The Sirius, Coma, Hyades, Pleiades, Hercules, and HR 1614 moving groups are clearly matched with our over/under estimated regions and this shows that our GMM model is classifying structures on larger scales than individual moving groups, and could potentially be used to remove the underlying smooth component for better study of these groups. 

Figure \ref{fig:real5} shows the augmented Toomre diagram for the best-fit 5-component Gaussian mixture model of Gaia-APOGEE. The components identified are strikingly similar to those picked out by the best-fit model for the mock catalog (Figure \ref{fig:mock5}). This is borne out by examining the distributions of various other stellar properties in Figure \ref{fig:real5dists}: though (sadly) we cannot show the \dform values and ages for the real catalog, APOGEE provides magnesium-to-iron ratios, Gaia provides heights above the disk plane at present day, and the astroNN Value-Added Catalog provides ages and estimated orbital properties \citep{2019MNRAS.489.2079L,2019MNRAS.489..176M}. For this catalog, in addition to quality cuts that we apply on the observed and mock catalogs, we select the astroNN ages for stars at [Fe/H]$>-0.5$ because there is not any training set stars for the catalog with low metallicities. The age distributions panel in Figure \ref{fig:real5dists} is from the astroNN Value Added Catalog and it shows the same pattern as in the mock catalogs: an exclusively old halo component, a young thin disk, and three intermediate groups with distinct distributions. The rose and the black components are close to each other in age, but the orange component is much younger on average.

Figure \ref{fig:astroNN-eccen} shows the orbital eccentricity distributions of the real data, respectively, for subpopulations classified using our mixture model. From the orbital eccentricity distributions, we see that stars in the halo component have more elliptical orbits and thin disk stars are on circular orbits, also as expected. The three thick disk components in this space are also different from each other; the orange component contains orbits that are generally more circular than those in the black and rose components.

Examining the alpha abundance distributions bears out our intuition from the mock catalogs as well. The Milky Way famously has a bimodal distribution in $[\alpha/\textrm{Fe}]$ \citep[e.g.][]{nidever_2015, Hayden2015} and we see that the least alpha-enhanced stars are associated with the thin disk component, while the halo includes the most enhanced ones. The black and rose components include some stars with both alpha abundances while the orange component is mainly composed of stars with \emph{some} alpha-enhancement, but less than the other two. However, the black and rose components are not completely identical: the metallicity distribution for the black component extends to Solar metallicity and above while the rose component truncates at lower [Fe/H]. The black component also has a larger tail at high $V_{XZ}$, extending as far as the orange component does, while the rose component is somewhat less extended in random energy. Interestingly, the orange component has higher mean $V_Y$ than all but the thin disk, yet also the widest spread in $V_{XZ}$ of any but the halo component.

\section{Discussion}
\label{sec:discussion}

In this paper we derived a best-fit, 5-component Gaussian mixture model of the Solar neighborhood in augmented Toomre space (velocities plus iron abundances). Despite extremely limited assumptions, the components picked out by the model in both the mock and real Gaia-APOGEE catalogs generally reflect common interpretations of the origin of various kinematic subpopulations. Based on our parallel analysis of simulated data, the model appears to be flexible enough to accommodate both asymmetric drift and thick disk subpopulations with different histories. 
\begin{figure*}
    \centering
    \includegraphics[width=\textwidth]{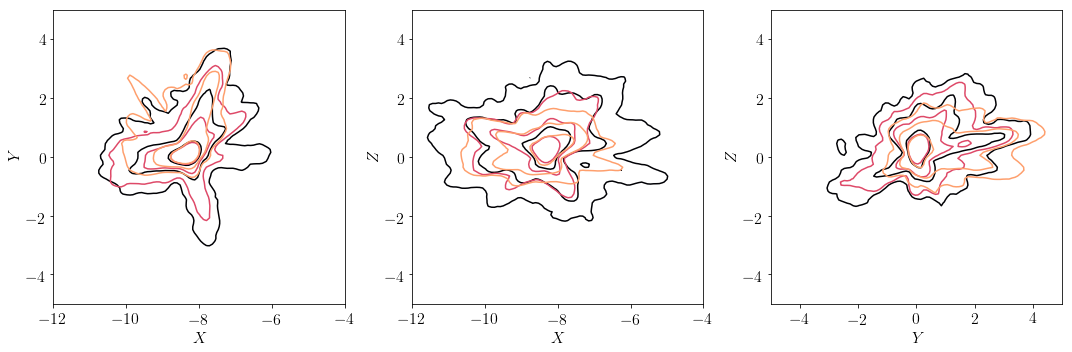}
    \includegraphics[width=\textwidth]{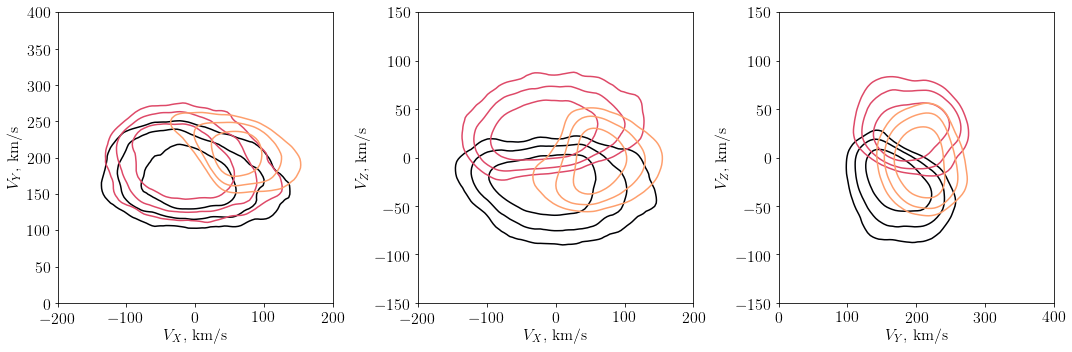}
    \caption{Spatial (top row) and velocity (bottom row) distributions of components I1, I2, and I3 in the mixture model of the real data. The black component (I3) contains stars with preferentially negative $V_Z$, while the rose component (I1) has mostly stars with positive $V_Z$. These two components comprise the velocity-asymmetric, alpha-enhanced thick disk, with one component used to represent each side of the ``wave.'' The most metal-rich thick-disk-like component (I2; orange) has very high random energy orbits nearly at the Solar circular velocity, and contains stars that have been heated by interactions with satellite galaxies such as Sagittarius.}
    \label{fig:xv-intermed}
\end{figure*}

\subsection{Origins of the families in our neighborhood}
In \S\ref{sec:real} we discuss the properties of the five components identified by our mixture model in the Gaia-APOGEE crossmatch. By analogy with similar distributions in our simulations, we propose the following scenario that is consistent with the properties of these different components and with other observations of the Galaxy.

\begin{itemize}
    \item[(a)] The \textbf{halo component} is extremely old yet has a relatively high mean metallicity, resembling the early history of \mm. This supports the idea that the majority of the halo component is from \textbf{one or two early, massive mergers} \citep[e.g.][]{2018MNRAS.478..611B, 2018Natur.563...85H, 2019MNRAS.488.1235M,2020arXiv200710374H,2020MNRAS.497..747S}.

\item[(b)] The most \textbf{metal-rich thick-disk-like component} (I2 in table \ref{tbl:realModel}; shown in orange) has very high random energy yet still orbits nearly at the Solar circular velocity, and appears to be made up mostly of stars with solar values of $[\alpha/\textrm{Fe}]$ or slightly higher with younger estimated ages. This resembles intermediate components in our \mf\ simulation that contain stars disturbed by a late merger. We propose that this component contains \textbf{stars that have been heated by interactions} with satellite galaxies such as Sagittarius \citep[e.g.][]{Villalobos2008, 2018ApJ...854...47S, 2018Natur.561..360A, 2018MNRAS.481..286L, 2018MNRAS.480..652E, ma2017}.

\item[(c, d)] The two \textbf{metal-poor thick-disk-like components} One component (I3, shown in black) has a random energy distribution as wide as I2 and extends to similarly high metallicity, but has lower orbital velocity and contains far more alpha-enhanced stars. It also seems to contain stars with preferentially negative $V_Z$, while the other component (I1, shown in rose) has mostly stars with positive $V_Z$ (Figure \ref{fig:xv-intermed}). I1 also has a slightly narrower $V_{XZ}$ distribution and slightly higher mean $V_Y$, but is otherwise elementally and spatially similar to I3. These two components comprise the \textbf{velocity-asymmetric, alpha-enhanced thick disk} \citep{2012ApJ...750L..41W}, with one component used to represent each side of the ``wave.'' Consistent with observations of the velocity asymmetry \citep[e.g. Figure 3 of][]{2012ApJ...750L..41W}, we find that the negative-velocity component (I3) has a wider velocity spread and includes stars both above and below the disk, while the positive-velocity component (I1) is more coherent in velocity, more spatially confined, and slightly prefers positive $z$ (Figure \ref{fig:xv-intermed}). In terms of formation mechanism, these components resemble the \textbf{radially mixed thick disks} seen in all three simulations, but given the MW's quiet recent accretion history, seem most closely to resemble the black component of the three \mi\ simulations.

\item[(e)] The \textbf{thin disk component} contains young, non-alpha-enhanced stars consistent with \textbf{recent star formation in a cold gas disk}, after a late influx of gas that reset the local $[\alpha/\textrm{Fe}]$ ratio \citep[e.g.][Wetzel et al. ~(in prep.)]{Mackereth2017a,2018MNRAS.477.5072M}.

\end{itemize}

Finally from the weights $\tau$ in Table \ref{tbl:realModel} we can estimate the proportion of stars with each of these origins in the APOGEE (i.e., evolved-star) view of our Solar neighborhood. According to our mixture, the APOGEE sample is \textbf{4\% halo (a), 22\% interaction-heated (b), 27\% radially mixed, asymmetric thick disk (c, d), and 46\% thin disk (e).} We caution that this breakdown is modulated by the APOGEE selection function; we plan to correct for this in future work using our new mock catalogs as a testbed.

\subsection{Further applications of the mixture model}
Our strategy offers a probabilistic approach for selecting stars that are likely to belong to a particular population, as an alternative to making hard cuts on the data, through analogy with state-of-the-art simulations of galaxies. The resulting model can be used either to classify and study the stars within the modeled dataset (in this case the Gaia-APOGEE catalog), or to predict the composition of other datasets that measure the same parameters. 

A trained Gaussian mixture model can be used to identify members of its components in any data set where the same features are available. To identify structural components of the Galaxy, we built a mixture model using 3D kinematics and metallicities of stars observed with Gaia-APOGEE, but this model can be used to probabilistically classify any star with a measured 3D velocity and [Fe/H]. Care should be taken to use the same metallicity scale when combining data from multiple sources. The ``validity volume'', where the GMM has been constructed, should be considered as well. An already-trained mixture model is especially useful for identifying members of Galactic components in smaller surveys, or ones which have a more complicated selection function, and are thus unlikely to independently constrain a mixture model. This is likely to be especially useful given that while Gaia provides all-sky coverage for proper motions and distances, radial velocities and elemental  abundances for most stars are determined by an ensemble of ground-based spectroscopic surveys, each with a different selection function, sky coverage, and target depth. Mixture modeling will thus supply a crucial tool to relate stars in the same population that have been observed by different instruments, unifying our chemodynamical view of the Galaxy.

\section*{Acknowledgments}
The authors thank Jeff Newman, the Milky Way as A Galaxy (MWAG) working group of SDSS-IV, and the Dynamics group at the Center for Computational Astrophysics, Flatiron Institute for valuable insights in preparing this paper.

FN acknowledges support from the National Science Foundation Graduate Research Fellowship under Grant No. DGE-1845298. 

RES acknowledges support from NASA grant 19-ATP19-0068, NSF grant AST-2009828, and HST-AR-15809 from the Space Telescope Science Institute (STScI), which is operated by AURA, Inc., under NASA contract NAS5-26555.

We performed this work in part at the Aspen Center for Physics, supported by NSF grant PHY-1607611, and at at KITP, supported by NSF grant PHY-1748958.

AW received support from NASA through ATP grants 80NSSC18K1097 and 80NSSC20K0513; HST grants GO-14734, AR-15057, AR-15809, and GO-15902 from STScI; a Scialog Award from the Heising-Simons Foundation; and a Hellman Fellowship.

JTM acknowledges support from the Banting Postdoctoral Fellowship programme administered by the Government of Canada, and a CITA/Dunlap Institute fellowship. The Dunlap Institute is funded through an endowment established by the David Dunlap family and the University of Toronto.

AB acknowledges support from NASA through HST grant HST-GO-15930.

SLM acknowledges support from the UNSW Scientia Fellowship program and the Australian Research Council through grant DP180101791.

CAFG was supported by NSF through grants AST-1715216 and CAREER award AST-1652522; by NASA through grant 17-ATP17-0067; by STScI through grant HST-AR-16124.001-A; and by the Research Corporation for Science Advancement through a Cottrell Scholar Award and a Scialog Award.

The authors thank the Flatiron Institute Scientific Computing Core for providing computing resources that made this research possible, and especially for their hard work facilitating remote access during the pandemic. Analysis for this paper was carried out on the Flatiron Institute's computing cluster \texttt{rusty}, which is supported by the Simons Foundation.

Simulations used in this work were run using XSEDE supported by NSF grant ACI-1548562, Blue Waters via allocation PRAC NSF.1713353 supported by the NSF, and NASA HEC Program through the NAS Division at Ames Research Center. 

This project was developed in part at the 2017 Heidelberg Gaia Sprint, hosted by the Max-Planck-Institut f\"{u}r Astronomie, Heidelberg.

This work has made use of data from the European Space Agency (ESA) mission Gaia (http://www.cosmos.esa.int/gaia), processed by the Gaia Data Processing and Analysis Consortium (DPAC, http://www.cosmos.esa.int/web/gaia/dpac/consortium). Funding for the DPAC has been provided by national institutions, in particular the institutions participating in the Gaia Multilateral Agreement. 

Funding for the Sloan Digital Sky Survey IV has been provided by the Alfred P. Sloan Foundation, the U.S. Department of Energy Office of Science, and the Participating Institutions. SDSS acknowledges support and resources from the Center for High-Performance Computing at the University of Utah. The SDSS web site is www.sdss.org.

SDSS is managed by the Astrophysical Research Consortium for the Participating Institutions of the SDSS Collaboration including the Brazilian Participation Group, the Carnegie Institution for Science, Carnegie Mellon University, the Chilean Participation Group, the French Participation Group, Harvard-Smithsonian Center for Astrophysics, Instituto de Astrof\'{i}sica de Canarias, The Johns Hopkins University, Kavli Institute for the Physics and Mathematics of the Universe (IPMU) / University of Tokyo, the Korean Participation Group, Lawrence Berkeley National Laboratory, Leibniz Institut f\"{u}r Astrophysik Potsdam (AIP), Max-Planck-Institut f\"{u}r Astronomie (MPIA Heidelberg), Max-Planck-Institut f\"{u}r Astrophysik (MPA Garching), Max-Planck-Institut f\"{u}r Extraterrestrische Physik (MPE), National Astronomical Observatories of China, New Mexico State University, New York University, University of Notre Dame, Observat\'{o}rio Nacional / MCTI, The Ohio State University, Pennsylvania State University, Shanghai Astronomical Observatory, United Kingdom Participation Group, Universidad Nacional Aut\'{o}noma de M\'{e}xico, University of Arizona, University of Colorado Boulder, University of Oxford, University of Portsmouth, University of Utah, University of Virginia, University of Washington, University of Wisconsin, Vanderbilt University, and Yale University.

\software{Numpy, matplotlib, scikit-learn, astropy, gizmo--analysis, halo--analysis}

\bibliographystyle{aasjournal}
\bibliography{references, apo2_refs}


\end{document}